\documentclass{sig-alternate}

\usepackage[svgnames, x11names]{xcolor}

\usepackage[normalem]{ulem} % required for strikeout font

\usepackage[latin]{babel}
\usepackage{blindtext}

\usepackage{listings}
\lstdefinelanguage{XML}
{
basicstyle=\ttfamily\footnotesize,
  morestring=[b]",
  moredelim=[s][\bfseries\color{Maroon}]{<}{\ },
  moredelim=[s][\bfseries\color{Maroon}]{</}{>},
  moredelim=[l][\bfseries\color{Maroon}]{/>},
  moredelim=[l][\bfseries\color{Maroon}]{>},
  morecomment=[s]{<?}{?>},
  morecomment=[s]{<!--}{-->},
  commentstyle=\color{gray},
  stringstyle=\color{blue},
  identifierstyle=\color{red}
}
\graphicspath{./figures/}
\usepackage[caption=false,font=normalsize,labelfont=sf,textfont=sf]{subfig}

\usepackage{algorithmicx}
\usepackage{algpseudocode}
\usepackage[ruled]{algorithm}
\definecolor{light-gray}{gray}{0.75}
\algrenewcommand{\algorithmiccomment}[1]{\hskip3em{{\footnotesize \textcolor{light-gray}{$\blacktriangleright$}}} \emph{\color{gray} #1}}

\usepackage{multirow}
\usepackage{booktabs}
\usepackage{hyperref}

\usepackage{xspace}
\usepackage[nocompress]{cite}

\usepackage{enumitem}

\usepackage{mathtools}
\DeclarePairedDelimiter{\ceil}{\lceil}{\rceil}

\usepackage{mathtools}

\begin{document}
%
% --- Author Metadata here ---
 \conferenceinfo{Under review for conference}{}
%\CopyrightYear{2007} % Allows default copyright year (20XX) to be over-ridden - IF NEED BE.
%\crdata{0-12345-67-8/90/01}  % Allows default copyright data (0-89791-88-6/97/05) to be over-ridden - IF NEED BE.
% --- End of Author Metadata ---

\title{Elastic Resource Allocation for Distributed Graph Processing Platforms}

\numberofauthors{1} %  in this sample file, there are a *total*
\author{
% 1st. author
\alignauthor
Ravikant Dindokar and Yogesh Simmhan\\
       \affaddr{Supercomputer Education and Research Centre}\\
       \affaddr{Indian Institute of Science}\\
       \affaddr{Bangalore, India}\\
       \email{ravikant7@ssl.serc.iisc.in, simmhan@serc.iisc.in}
}
% There's nothing stopping you putting the seventh, eighth, etc.
% author on the opening page (as the 'third row') but we ask,
% for aesthetic reasons that you place these 'additional authors'
% in the \additional authors block, viz.
% \additionalauthors{Additional authors: John Smith (The Th{\o}rv{\"a}ld Group,
% email: {\texttt{jsmith@affiliation.org}}) and Julius P.~Kumquat
% (The Kumquat Consortium, email: {\texttt{jpkumquat@consortium.net}}).}
% \date{30 July 1999}
% Just remember to make sure that the TOTAL number of authors
% is the number that will appear on the first page PLUS the
% number that will appear in the \additionalauthors section.

\maketitle
\begin{abstract}
Distributed graph platforms like Pregel have used vertex-centric programming models to process the growing corpus of graph datasets using commodity clusters. The irregular structure of graphs cause load imbalances across machines operating on graph partitions, and this is exacerbated for non-stationary graph algorithms such as traversals, where not all parts of the graph are active at the same time. As a result, such graph platforms, even as they scale, do not make efficient use of distributed resources. Clouds offer elastic virtual machines (VMs) that can be leveraged to improve the resource utilization for such platforms and hence reduce the monetary cost for their execution. In this paper, we propose strategies for \emph{elastic placement of graph partitions on Cloud VMs} for a subgraph-centric programming model to reduce the cost of execution compared to a static placement, even as we minimize the increase in makespan. These strategies are innovative in modeling the graph algorithm's behavior \emph{a priori} using a metagraph sketch for the large graph. We validate our strategies for several graphs, using runtime traces for their distributed execution of a Breadth First Search (BFS) algorithms on our subgraph-centric GoFFish graph platform. Our strategies are able to reduce the cost of execution by up to $42$\%, compared to a static placement, while achieving a makespan that is within $29$\% of the optimal.
\end{abstract}

%A category including the fourth, optional field follows...
%\category{D.2.8}{Software Engineering}{Metrics}[complexity measures, performance measures]

\category{}{Networks}{Cloud computing}
\category{}{Computing methodologies}{Distributed computing methodologies}
\category{}{Computing methodologies}{Planning and scheduling}

\terms{Design, Performance, Experimentation}


\section{Introduction}
% \ysnote{Motivation for graph processing}
% \ysnote{distributed graph platforms, Clouds}
% electric and transportation networks, genomics and connectomics, as well as the more common social and web graphs
% GraphLab, Pregel/Giraph, Giraph++/Blogel/GoFFish. Use of Clouds. Distinct from HPC.
% Discuss what a superstep is.
Graph algorithms are challenging to design, program and execute in parallel due to their irregular nature. There has been a rapid growth of large graph datasets, ranging from social networks~\cite{facebook-trillion} and knowledge graphs~\cite{nell}, to power and road infrastructure graphs~\footnote{http://www.dis.uniroma1.it/challenge9/download.shtml} and connectivity of the Internet of Things. At the same time, there has been a push toward developing Big Data platforms for graph processing on commodity clusters and Clouds. Distributed graph programming models such as Google's Pregel~\cite{pregel} and GraphLab~\cite{graphlab} leverage an intuitive vertex-centric approach to specifying graph algorithms, where users specify the logic for a single vertex and this is executed in parallel across all vertices, with message passing or state transfer between them. These have been extended to other component-centric variants~\cite{giraph++,goffish} that execute iteratively using a Bulk Synchronous Parallel (BSP) model. However, despite their use of commodity clusters, there has not been work on effectively leveraging \emph{elastic} Cloud resources for such irregular graph platforms.

%\ysnote{Effectively using elastic resources for irregular applications is challenging}

% Gaps in scalable and efficient execution
% Partiitoning has tried to address irregular structure
% But non-stationary structure still a challenge
Many of the newer graph platforms have been able to scale with the size of the graph, but they do not necessarily achieve a high efficiency of execution. For e.g., a vertex-centric programming model like Pregel is nominally able to achieve a balanced computation across different partitions (machines) due to similar number of vertices on each partition~\cite{pregel, giraph}, and edge-balanced partitioning has been attempted too~\cite{gps}. But a balancing of the topology across machines translates to a balanced CPU utilization of all the machines only for \emph{stationary graph algorithms}~\cite{mizan}, where all vertices or edges are actively computing during the entire algorithm. Such algorithms, like PageRank or Bi-partite connectivity, can achieve good load balancing across machines. 

However, when the algorithm itself moves to different parts of the graphs over different supersteps (iterations), the load across different machines gets out of balance. Such \emph{non-stationary algorithms}~\cite{mizan} include traversals (e.g., breadth-first search, single source shortest path) and centrality (e.g. between-centrality) algorithms. For e.g., in BFS, the partition containing the source vertex is active in the first superstep and as the traversal progresses, its neighboring partitions get active in subsequent supersteps, and so on. % When the graph has been partitioned based on the topology, i.e., by keeping tightly connected vertices in the same partition, we can see that the set of frontier (i.e. active) partitions change for different supersteps. 
As a result, only a subset of the partitions are active at a time with their host machine's CPU being used, even as the other machines holding inactive vertices are under-/un-used. For example, the average utilization for BFS, using a subgraph-centric model, over the USA Road Network ($23$M vertices, $85$M edges) running on $8$ machines is $35\%$ (Fig.~\ref{fig:lowutil:usrn16}).% \ysnote{RAVI: Plot for the CPU utilization or partitions/vertices active across supersteps.}\drnote{done}

% Goal
Cloud computing has been actively used by Big Data platforms due to it easy access to commodity infrastructure. One of its key benefits is the elastic access to resources, that allows virtual machines (VMs) to be acquired and released on-demand with a pay-as-you-go pricing. Infrastructure as a Cloud (IaaS) rent out VMs by the hour (Amazon AWS) or even by the minute (Google Compute and Microsoft Azure). Irregular distributed graph algorithms, operating on large graphs, can take $100$'s of core-minutes to run on commodity hardware. \emph{Hence, this offers the opportunity to control the VM elasticity such that only partitions that are active take up VM resources}, thereby reducing the monetary cost of execution. There has been limited work on actively using Cloud elasticity for graph platforms. We address this gap in this paper by proposing partition activation and placement strategies on Clouds for graph platforms.

% Approach / Contributions
% De-couple partitioning from the placement strategy.
Two key intuitions drive our approach: (1) We decouple partitioning of the graph from their placement on VMs for executing a particular algorithm; and (2) We utilize a metagraph~\cite{parlearning} sketch of the whole graph to \emph{a priori} model the progress of the algorithm onto various partitions, which guides our placement strategy at each superstep. As a result, for each superstep, we are able to activate VMs and place relevant partitions on them, and conversely, deactivate VMs without active partitions. In the process, we reduce the monetary cost of execution on the Cloud with minimal impact on the runtime of the algorithm.

% Organization
The rest of this paper is organized as follows. In \S~\ref{sec:related} we discuss related work on distributed graph processing and partitioning strategies. In \S~\ref{sec:background}, we review our prior work on the GoFFish subgraph-centric distributed platform and the concept of metagraphs, which are used in our approach and evaluation. We define the system model considered and formalize the problem of partition placement onto VMs in \S~\ref{sec:problem}. \S~\ref{sec:elastic} introduces our proposed partition placement and VM activation strategies. We validate these strategies in \S~\ref{sec:evaluation} using execution traces collected from real graphs for the Breadth First Search (BFS) algorithm and evaluate the potential benefits in reducing the VM cost. Lastly, in \S~\ref{sec:conclusion} we present our conclusions and discuss future work.

\section{Related Work}
\label{sec:related}

%%%%
%\ysnote{Distributed Graph Processing, partitioning, dynamic migration (GPS, Mizran), Non-stationary algorithms}\\
%\drnote{Should Goffish especially the arxivx and blogel partitioning need to be introduced here?}\\
% distributed graph paltforms, Pregel
MapReduce has been a staple platform for Big Data processing but graphs present challenges to its tuple-centric model. Its shortcomings for graph algorithms, which tend to be iterative, are due to the repetitive costs for disk I/O, both to reload the graph and to pass state, for each iteration~\cite{cise-mapreduce-graph}. To address this, Google's Pregel~\cite{pregel} offers a vertex-centric programming model that keeps the graph in memory, and uses an iterative Bulk Synchronous Parallel (BSP) execution model where messages are passed between vertices for state transfer at superstep boundaries. % over the MapReduce model. Pregel is built on top of the Valiant BSP processing model\cite{•}. Pregel provides simple vertex centric API in which logic is specified from the perspective of a single vertex. The execution takes places as a series of iterations called superstep, separated by synchronization barrier.
Within a superstep, active vertices execute their \texttt{compute()} method once, in parallel, and this method has access to that vertex's prior state and any incoming messages from its neighbors from the previous superstep. %These lead to significant performance improvements over MapReduce and allow vertex-level parallelism. % . Vertices communicate using message passing, and the messages are delivered at the end of a superstep, and available to the compute method in the next superstep. 

%%%%
% active and inactive vertices
All vertices are \emph{active} when the Pregel application begins, and a vertex can locally \emph{vote to halt} as part of its compute logic, when it ``appears'' to have finished. This makes it \emph{inactive}, and the compute method of an inactive vertex is not invoked in a superstep. Inactive vertices can be revived and active again when a new message is received by them at a superstep. When all vertices are inactive in a superstep, a global \emph{vote to halt} is reached and the application terminates.

%%%%
%\ysnote{other graph platforms. include goffish, giraph++ and blogel. talk about similaritities with pregel.}
Pregel has spawned Apache Giraph~\cite{giraph} as an open source implementation, and other optimizations to its programming and execution models. Giraph++~\cite{giraph++}, Blogel~\cite{blogel} and our own work on GoFFish~\cite{goffish} coarsen the programming model to operate on partitions or subgraphs, with Giraph++ using partitions, GoFFish on subgraphs (weakly connected components, WCC) and Blogel on either vertices or blocks (WCC). This gives users more flexible access to graph components that can lead to faster convergence, and also reduces fine-grained vertex-level communication. This paper aims to use elastic Cloud VMs for such component-centric systems.

Distributed graph processing systems divide the graph into a number of partitions which are placed across machines for execution. The quality of partitioning impacts the load on a machine, cost of communication between vertices, and the iterations required to converge. % when vertex from one node wants to communicate with vertex in other node. This depicts partitioning has great impact on runtime of the overall system and implies the need for partitioning to be done in cleaver way.
A variety of partitioning techniques have been tried. Giraph's default partitioner hashes vertex IDs to machines, to balance the number vertices per machine. Other approaches that balance the number edges per partition, for algorithms that are edge-bound, have been tried~\cite{edgebasedpartitioning}. GoFFish tries to balance the number of vertices per partition while also minimizing the edge cuts between partitions. This gives partitions with well-connected components that suits its subgraph-centric model. Multi-level partitioning schemes have also been identified to improve the CPU utilization~\cite{goffish-arxiv}. Blogel further uses special 2D partitoners for spatial graphs to improve the convergence time for reachability algorithms. 

However, unlike stationary algorithms~\cite{mizan} like PageRank where all vertices are active on all supersteps, non-stationary traversal algorithms like BFS have a varying frontier set of vertices that are active in each iteration. This results in an uneven workload across different machines. \emph{Hence, a single partitioning scheme, however good its quality may be in achieving some topological balancing, cannot offer compute balancing across hosts, minimize communication and ensure fast convergence, for all types of graphs and algorithms.} We address the lack of compute balancing for non-stationary algorithms here, and the associated suboptimal Cloud costs.

Platforms like Mizan~\cite{mizan} %go beyond the initial static partitioning % . Mizan does not make any assumption about initial graph partitioning and runtime characteristics of algorithm. Mizan has a module called
partially address this imbalance by performing vertex migration based on the number of outgoing and incoming messages to vertices and the
execution time of a vertex in a superstep. This identifies overloaded machines at the end of each superstep, and the vertices to migrate to less loaded machines. However, this runtime decision causes additional coordination costs to decide and move vertices, and re-synchronize before the next superstep's compute can be started. % The metric used to make migration decisions may also fail when \ysnote{RAVI: a brief comment here}. 
This can result in an increased makespan for non-stationary algorithms, which their results do not document, and also affect its correctness~\cite{mizan-incorrect}.% execution is also like traversal where the imbalance (and hence benefit) is likely to be high. %Migration of vertices (ownership + state) is an expensive operation. Profiling of migration cost for non stationary algorithm is not mentioned in the work\cite{•}. So it is unclear whether this appraoch will be effective for non stationary algorithms or not. \\
%%%%

%%%%
%\ysnote{Other schemes to mitigate impact of static partitioning. What are their gaps relative to out contribution?}
GPS~\cite{gps} too adopts dynamic re-partitioning to reduce communication by co-locating vertices that communicate beyond threshold onto the same machine while also balancing the number of vertices per machine. It only takes into account the outgoing messages from a vertex for this decision, which is in-sufficient for load balancing in non-stationary algorithms.  %if some vertex from worker $Wi$ is observed to send/receive message from worker $Wj$, above specific threshould, then vertex can be reassigned to worker $Wj$, with guarantee that the number of vertices in each worker does not change. It also distributes the adjacency list to all compute nodes to achieve higher performance. The drawback of this approach is that 
Similarly, \cite{wind} tries to balance the workload across the machines by an experimental study of the graph algorithm and a prediction of the number of active vertices in the next superstep to perform vertex migrations. Such analytical and experimental predictions are difficult and costly for new graphs or algorithms. % has studied behaviour of graph algorithms in terms of active vertices termed as window. In this work they have estimated next working window from the current one and have come up with vertex migration plan to achieve the load balancing. They have implemented their prototyped approach on HAMA. Their approach requires prediction of working window which is difficult problem. They rely on patterns observed through extensive experimental studies.   

Two key distinctions between these works on vertex migration and ours is, (1) rather than balance the workload across a \emph{static set of machines}, we scale-out or -in the number of \emph{elastic VMs} to match the workload on a superstep, and (2) we use an \emph{a priori} analysis of the graph algorithm on a coarse metagraph to model the execution time of partitions \emph{rapidly}, and use this to decide both the number of VMs required and the placement of partitions on VMs. This static planning of migration and scaling for each superstep helps hide the migration costs by interleaving data movement with compute, and can avoid increasing the makespan.

\emph{To our knowledge, there is no detailed, existing work on the effective use of elastic VMs to execute component-centric graph frameworks.} Our prior work \cite{redekopp:ipdps:2012} briefly examines dynamic scaling of BSP workers on elastic VMs for a vertex-centric model to reduce the cost of execution of the Betweenness Centrality (BC) algorithm. This uses an intuition that BC has a sinusoidal number of active vertices when launching traversals from multiple source vertices, and this can be used to control the number of VMs. % This was briefly discussed using an analytical calculation and works only for BC.
We generalize this model in the current paper using the notion of metagraphs than can be used for many graph algorithms, including BC, and without needing to empirically observe the algorithm's execution on the whole graph. %shown when 50\% of vertices are active for the current set of traversals. More sophisticated approach is required to fully utilize the elasticity of cloud, to get performance comparable to BSP model along with significantly lower cost.

There has also been some work on algorithmic analysis of component-centric graph algorithms to guide their efficient execution~\cite{PPA, widom:sigmod}. These identify desirable properties for algorithms designed for distributed graph processing systems like Pregel. Our work is complementary to such algorithmic innovations, and such techniques can be applied to analyzing the metagraph too. %\cite{PPA paper} identifies desirable properties for algorithms designed for distributed graph processing systems like Pregel, which include linear space, communication and compute cost per iteration and logarithmic number of iterations. \ysnote{widom's SIGMOD/PVLDB paper on spanning tree also has algorithmic insights?}%They have designed algorithms for CC, BCC and SCC problems to satisfy above mentioned properties and observed good performance improvements. 
%But developing efficient algorithms, is a difficult task.  

%%%%
Besides Pregel-like systems there are several other distributed graph processing systems. GraphLab~\cite{graphlab} uses an asynchronous, though vertex-centric, programming model where a vertex can directly access its neighboring vertex and edge values, without the need for messaging. GraphLab also adopts a % two level partitioning method that minimizes the edges among graph partitions and allows for a 
fast repartitioning scheme that is user-driven. % Asynchronous execution has to deal with  additional complexity of data locking and unlocking and its associated cost.  
Powergraph~\cite{powergraph} follows a vertex based partitioning and replication to distribute load based on edges. %The paper\cite{} has evaluation metrics defined for stationary algorithms only. So it is unclear, how the system will perform for non stationary algorithms and  graphs which do not follow power law distribution.  \\ 
Sedge\cite{sedge}, is a distributed graph querying system over a set of non overlapping partitions, and it can reduce the communication costs at runtime by creating new partitions or replicating them. Trinity~\cite{trinity} is a distributed in-memory key-value store to perform graph analytics such as path traversal on RDF data, which is treated as a graph. However, none of these distributed graph processing systems are optimized for elastic Cloud resources, which is the emphasis of this paper. % them across different machines for improved performance. But the domain for the workload is multiple concurrent queries which is orthogonal to  graph algorithms space.

%GPS\cite{•} \ysnote{graphlab precedes pregel} and GraphX\cite{•}.  

%%%%

%%%%

% \ysnote{distinguish between static/runtime partitioning for a statis set of machines, and partitioning for an elastic set of machines. Why is our work important. What have other done for elastic resources?}

%%%%

%%%%

% All the approaches discussed before, primarily work on fixed number of machines. \ysnote{I thought one of the elastic paper was using different number of VMs?} That is all worker machines remain active throught lifecycle of graph algorithm and an attempt is made to balance the work among these machines. 
%%%%

% In this paper, our goal would be to use the elastic property of cloud systems and accomodate different partitions of the graph in minimum number of machines, without disturbing the BSP model of the graph processing, with minimum VM cost.

\section{Background}
\label{sec:background}
We use the GoFFish subgraph-centric graph programming model to translate and validate our approach of using elastic placement strategies for non-stationary graph algorithms. As background, we give details on GoFFish's graph data distribution and execution model, and the ability to construct a metagraph on top of large graphs to analyze the execution of graph algorithms.

\subsection{GoFFish Subgraph-centric Model}
As described before, GoFFish~\cite{goffish} is a distributed graph processing framework that follows a BSP model to execute graph application written using a subgraph-centric programming model. %Users implement a \texttt{compute()} method for an entire \emph{subgraph}, that are weakly connected component in a partition. 

\begin{figure}[t]
\centering%~%
\includegraphics[width=1\columnwidth]{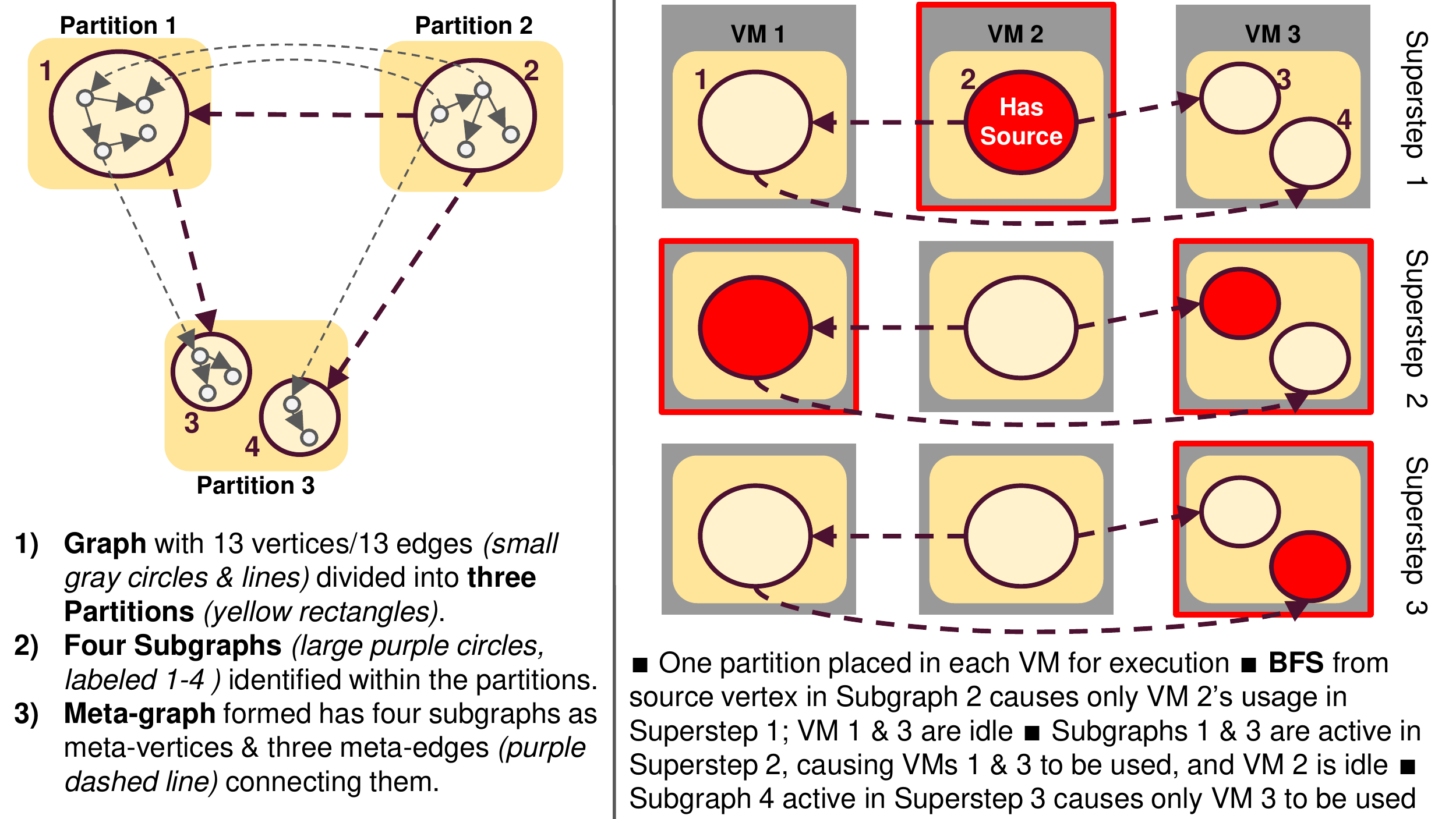}
\caption{Metagraph formation from partitioning graph (left), and active subgraphs, along with VMs used, during BFS execution in GoFFish (right). }
    \label{fig:metagraph}
\end{figure}

Consider a directed graph $G=\langle V, E \rangle$ is partitioned into $m$ partitions using some relevant partitioning scheme that reduces the edge cuts across partitions and balances vertices in each, and these are distributed over $n$ machines, such that $m \geq n$. A single machine can hold one or more partitions. Partitions are vertex disjoint, meaning that a vertex belongs to exactly one partition. Edges connecting vertices in different partitions are termed as \emph{remote edges}, whereas those having both source and sink vertices in the same partition are called \emph{local edges}. % So if a graph $G$ is partitioned into $P_1, P_2, \ldots, P_m$ partitions, and for partition $P_i$, let $\langle V_i, E_i \rangle$  denote the corresponding vertices and edges then  $\bigcup _{i=1}^m V_i ~=~ V$ and  $V_i \bigcap V_j ~=~ \varnothing $ for all $i,j \in (1,m)$ and $i\neq j$.
Within each partition, \emph{subgraphs} are formed by identifying weakly connected components, meaning that a partition can have one or more subgraphs, and no two vertices belonging to two different subgraphs are connected by a local edge. % Let a partition $P_i$ contain subgraphs $SG_1, SG_2,  \ldots, SG_p$ such that $SG_k=\langle v_k, e_k, r_k \rangle$, where $v_k$, $e_k$ and $r_k$ denote the sets of local vertices, local edges and remote edges connecting to a vertex in a subgraph on a different partition, respectively. Then  $\bigcup _{k=1}^n v_k ~=~ V_i$ and $\bigcup _{k=1}^n e_k + \bigcup _{k=1}^n r_k ~=~ E_i$ for all $i,j \in (1,n)$. Let an edge $e_i$ is denoted by pair\big\{u, v\big\}, where u is source and v is sink vertex, then $ E_k\big\{V_i, V_j\big\} ~=~ \varnothing $ for all $V_i \in SG_i  , V_j \in SG_j , SG_i, such that SG_j \in P_i$ and $i\neq j$.\\
For e.g., Fig.~\ref{fig:metagraph} (left) shows a graph with 13 vertices and edges partitioned into three, with 4 subgraphs in all identified.

The \texttt{compute()} method defined by the user is applied to each subgraph in a superstep, and the method has access to all vertices, and local and remote edges in the subgraph. The method can update the state of local and remote edges, and pass messages to neighboring vertices or subgraphs connected through remote edges that are delivered at synchronized superstep boundaries. For e.g., Fig.~\ref{fig:metagraph} (right) shows a BFS that starts at a vertex present in subgraph 2, whose vertices are traversed in superstep 1, followed by traversal of vertices in neighboring subgraphs 1 and 3 in superstep 2, and their neighbor, subgraph 4, in superstep 3. A subgraph may be revisited too, or only a subset of its vertices be traversed, depending on the algorithm. The subgraphs can vote to halt and the execution stops when all subgraphs have voted to halt and no messages are in-flight. This model can also trivially implement a vertex-centric program.

%\ysnote{have a para describing the execution model, where the compute method of each subgraph is executed once per superstep, as long as it has not voted to halt or has not received any new message.  Maybe use a single figure to describe both metagraphs and subgraph-centric execution.} \ysnote{Also note that a vertex centric model can be implemented using this SG centric model too, so we're a superset of the vertex centric model.}

%\ysnote{no need to discuss adv since we've already stated it in related work. This section should be just statinh what is present, not why.} \delc{The subgraph centric model leads to several advantages over vertex cetric model. The messaging takes place between subgraphs across different partitions rather than vertices which may be within same partition in case of Pregel. The unit of computation, i.e. a subgraph is more coarser leading to faster convergenece for many graph algorithms.}

%\drnote{sholud we describe the partitioning techniques: flat and hierachiecal here?}\ysnote{no, not needed.}

\subsection{Metagraphs for Algorithm Modeling}
In our previous work~\cite{parlearning}, we have introduced the concept of \emph{metagraph} which is a coarse-grained sketch over large graphs, and is naturally suited to analyze subgraph-oriented graph applications. In a metagraph, each \emph{meta-vertex} is a disjoint \emph{subgraph} (connected component) in the original graph, and \emph{meta-edges} indicate remote edges that connect these subgraphs. The meta-vertices can have attributes like the number of local vertices and local edges that the subgraph has, and the weight of the meta-edge can indicate the number of remote edges between the subgraphs. Fig.~\ref{fig:metagraph}~(left) shows a metagraph with 4 meta-vertices and 3 meta-edges.

Depending on the type of graph and the partitioning scheme, metagraphs have meta-vertices and meta-edges that number several orders of magnitude smaller than the original graph. For e.g., we see metagraphs with 10's or 100's of meta-vertices for graphs with millions of vertices~\cite{parlearning}. As a result, this coarse-grained approximation of the large graph helps us rapidly analyze the behavior of several traversal algorithms, and can guide runtime operations. 

%Single Source Shortest Path (SSSP) on an unweighted graph
For e.g., when performing a BFS using a subgraph-centric model, our prior work~\cite{parlearning} showed that the order in which subgraph are potentially visited in each superstep can be determined. This is done by performing a BFS from the meta-vertex (subgraph) that holds the source vertex of the BFS, and traversing to neighboring meta-vertices. Each traversal is a superstep, and the meta-vertices visited corresponds to the subgraph(s) that will be active in that particular superstep. This is illustrated in Fig.~\ref{fig:metagraph}~(right).

This information, combined with an analytical model of the cost of a local BFS on a single subgraph, helped us estimate the number of supersteps required for the algorithm to converge. 
%\ysnote{RAVI: figure to explain both subgraph centric and meta graphs will help}\drnote{figure shared over drive }
The metagraph is simple to construct, either at graph partitioning time or by running a simple traversal algorithm using GoFFish that take a few seconds for even large graphs with millions of vertices. Also, the size of the metagraph itself is small enough that it can be analyzed on a single machine using a sequential algorithm.

In this paper, we reuse two important results from our previous work~\cite{parlearning} that supports our assumption of \emph{a priori} knowledge of the algorithm behavior. First, given the subgraph holding the source vertex and the metagraph, we can accurately determine the superstep at which a subgraph will be visited for the first time by a BFS and estimate the cost for local BFS to be performed on it. Second, we can predict the supersteps at which a subgraph may be revisited and the BFS potentially repeated on it. %This result will enable to estimate which subgraph will be active in which superstep of the SSSP algorithm execution.\drnote{should we talk about cost of generating a metagraph here i.e.  The time and space required to generate metagraphs (using a GoFFish job) is very small -- O(seconds) to generate, O(KB) to store.}

\section{Problem}
\label{sec:problem}
We discuss the high level problem and the system model as context, before giving a formal definition of our problem.
%\ysnote{We de-couple partitioning from the placement strategy. Give an intuition with fig on what this means.}
\subsection{Decoupling Partitioning from Placement}
Vertex and block-centric distributed graph platforms partition a large graph into many partitions, and workers in each host operate upon one or more partitions. These three steps -- that of \emph{partitioning} a graph, \emph{placement} of partitions onto a host, and \emph{assigning} workers (threads/processes) on a machine to operate on partitions present on it -- are loosely coupled decisions. For e.g., Giraph by default hashes a graph into as many partitions as requested by an application at runtime. There are one or more workers on each machine (equal to the number of mapper slots) and one or more partitions are assigned to a worker, and pushed to its machine. % the number of mappers, allowing multiple mappers per machine, and allows multiple threads per mapper to operate on the vertices in a partition, independently.}\drnote{ This is incorrect. The number of workers is equal to the number of mappers. The number of workers can be specified by user. Each worker can process more than one partition. The master determines how many partitions the graph will have, and assigns one or more partitions to each worker.  The number of partitions can also be specified by user using option  giraph.userPartitionCount. But the statement that these three steps are loosely coupled is correct since we can change the number of partitions and partitioning function in Giraph . About mapping to worker I am not sure. }
In GoFFish, we allow the user to specify the number of partitions to create at graph load time, allow multiple partitions per machine -- typically one per core, and allocate two threads per core to operate on the partition, and each thread works on one subgraph at a time within the partition. However, in both cases, once these bindings are made, they are retained for all supersteps.

\begin{figure}[t]
\centering%~%
\includegraphics[width=0.40\textwidth]{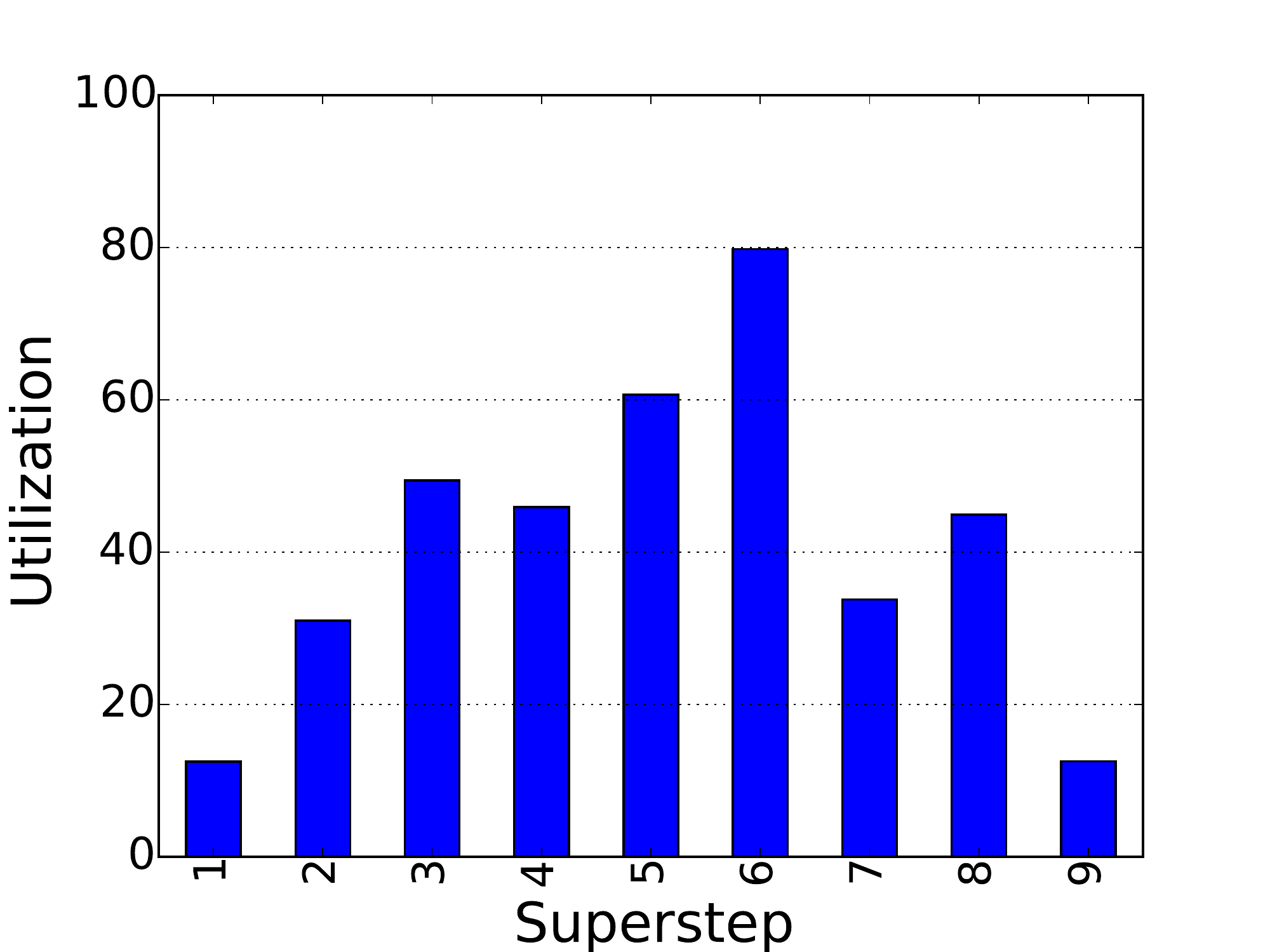}
\caption{CPU utilization \% over different supersteps, for US Road Network graph with 8 partitions executing BFS across 8 cores using GoFFish.}

    \label{fig:lowutil:usrn16}
\end{figure}

In each BSP superstep, vertices/subgraphs in all or some partitions may be actively performing computation. % , depending on whether a stationary or non-stationary graph algorithm is running, respectively. 
For non-stationary algorithms, such as BFS or SSSP, the inactive partitions, while taking up disk (and possibly memory) space, do not contribute to the computational usage on these machines. Fig.~\ref{fig:lowutil:usrn16} shows the average utilization of the hosts in each superstep when performing a BFS on a USRN graph (Table~\ref{tbl:datasets}) with 8 partitions, each using one core, on GoFFish. We see the CPU usage grow from 10\% to 80\% and drop again, depending on the number of active subgraphs. %, and consequently, their workers (hence cores) are idle while still being assigned to that partition.}    

Our focus here will be on such non-stationary graph algorithms. Since the compute load on a machine varies with each superstep as a result of which partitions on them are active or not, it can be efficient to consolidate the computational load of active partitions at each superstep on fewer (virtual) machines in order to increase the utilization of these machines. In the context of Cloud Virtual Machines (VMs), this also allows the VMs without any active partitions to be shutdown, and stop incurring monetary costs. This is the thesis of this paper, to explore \emph{how we can reduce the overall monetary cost for running the graph algorithm with minimal impact on the makespan of the algorithm, using partition placement strategies on elastic VMs based on their activation schedule across supersteps, as compared to a traditional hashing of partitions onto a static set of VMs.}

\subsection{System Model and Assumptions}
There are several reasonable assumptions that have to be satisfied in order for the above goal from being achieved. %We list these assumptions and justify they are reasonable.
\begin{itemize}[noitemsep,leftmargin=0.15in] %[noitemsep,topsep=0pt,parsep=0pt,partopsep=0pt]
\item \emph{VM costing should be at a fine time granularity.} Since the graph algorithms may run from 10's of seconds to 10's of minutes, elastic management of VMs will work only when VMs themselves can be acquired, released and billed at similar time scales. Both Google Compute and Microsoft Azure VMs allow billing at $1$~VM-min increments, with the former requiring a minimum billing of $10$~VM-mins. We use $1$~core-min as the smallest billing increment and unit of acquisition/release of small (1-core) VMs. We also assume that the startup and shutdown times are sufficiently small to not impact the costing or usage time of VMs. In future, with light-weight containerization of IaaS Clouds, even per-second billing may be possible.
% \item The superstep time has to be non-trivial, at least for the granularity of costing.
\item \emph{Framework support for elasticity.} The distributed graph platform should allow the persistence of state for a partition and its migration to different VMs at runtime, at superstep boundaries. Giraph support check-pointing of messages and state at supersteps. Other vertex-centric frameworks have also demonstrated this~\cite{mizan,gps}. GoFFish does not yet support this, though there is no design limitation that prevents this. We also investigate placement strategies that pin a partition to a single VM to avoid the need for migration.
%5) Startup and shutdown time should be small.
%\item The in-memory state should be persistable for movement.
\item \emph{Modeling graph algorithms.} The (non-stationary) graph algorithm itself should exhibit variable activation of partitions at different supersteps. Stationary algorithms, where all partitions are active at all supersteps, are unlikely to benefit. Also, we should be able to predict the activations of partitions for \emph{a priori} planning at launch time rather than runtime to avoid VM startup and data movement time from increasing the makespan, i.e. through prior planning, we can hide these by interleaving them with the compute time of a superstep. The meta-graph approach~\cite{parlearning} discussed above can enable this prediction. %next section discusses means to predict these times.
\end{itemize}

\subsection{Problem Definition}
Let a directed graph $\mathbb{G}=\langle V, E \rangle$ be partitioned into $n$ partitions, $P_1,P_2, \ldots,P_n$. Say we are given a graph application composed in a partition-centric manner, that takes $m$ supersteps to execute. Let there be a \emph{time function}, 
\[ \mathcal{A}:P_i \times s \rightarrow \tau_i^s \] 
that gives the computing time $\tau_i^s$ taken to execute a partition $P_i$ in a superstep $s \le m$ by a single exclusive virtual machine (VM). Here, if the function maps to a time value of $0$, that partition is not active in that superstep.

The \emph{minimum makespan} for this application is given by, 
\[ T_{Min} = \sum_{s=1}^m \tau_{Max}^s \text{~, where~~~} \tau_{Max}^s = \max_{i=1}^n\big( \tau_i^s \big)\] 
i.e., the minimum time to execute this application is the sum across all supersteps of the maximum computing taken among all partitions in each superstep. %, since the partition is the smaller unit of computing in a partition-centric model.

Say there is a \emph{partition placement function}, \[ \mathcal{M}:P_i \times s \rightarrow \upsilon_j \] that maps each partition $P_i$ for a superstep $s$ uniquely onto a virtual machine $\upsilon_j \in \upsilon_1, \upsilon_2, \ldots, \upsilon_l$ that can be used to execute that partition in a superstep. Similarly, let a partition presence function $\widehat{\mathcal{M}}(i,s,j)$ return $1$ if a partition $i$ is present in a VM $j$ in superstep $s$, and $0$ otherwise.

The \emph{actual makespan} for the application is given by summing up for each superstep, the time taken by the VM holding partitions that together take the longest time to execute sequentially in that superstep:
\[ T = \sum_{s=1}^m \Big( \max_{j=1}^{l}{ \big( \sum_{i=1}^n \widehat{\mathcal{M}}(i,s,j) \cdot \tau_i^s \big)} \Big) \]

If the cost for using a VM for a minimum time quanta $\delta$ is $\gamma$, the cost for this execution lies between $\Gamma_{Min}$ and $\Gamma_{Max}$,
\[ \Gamma_{Min} = \sum_{j=1}^l  \ceil[\Big]{ \frac{\sum_{s=1}^m \widehat{\mathcal{M}}(i,s,j) \cdot \tau_i^s}{\delta}} \cdot  \gamma\]
\[ \Gamma_{Max} = \sum_{s=1}^m \Big(\ceil[\Big]{\frac{\tau_{Max}^s}{\delta} } \cdot \mid \Upsilon_s \mid  \cdot  \gamma \Big) \]
where $\Upsilon_s$ is the set of active VMs in superstep $s$, and $\mid \Upsilon_s \mid$ is the number of active VMs in that set,
\[ \Upsilon_s = \{\upsilon_j \mid \widehat{\mathcal{M}}(i,s,j) = 1 \text{~for any~} 1 \le i \le n \} \]

$\Gamma_{Min}$ gives the minimum cost for all the VMs if each of their active times are summed over all supersteps and rounded up to the nearest $\delta$. This does not account for the cost of turning off and restarting VMs that are inactive in intermediate supersteps, which would causes their billing to be rounded up each time. Whereas, $\Gamma_{Max}$ gives the maximum cost required to run the application when each VM is billed in each superstep, independently, based on only the runtime of that superstep rounded up to the nearest $\delta$. The \emph{actual billing cost} for the application $\Gamma$ falls between $\Gamma_{Max}$ and $\Gamma_{Min}$. In practice, as we show, the actual value of $\Gamma$ depends on the strategy used to select and activate VMs when making the placement decision. %to reduce the cost, we would be using cost function as described in section $5.3$
%That is we are not considering the fact, a VM can be active in contiguous supersteps. 

\textbf{Problem:} Given a graph $\mathbb{G}$ and its partitions $P_i$, and the time function $\mathcal{A}$ for a graph application, the problem is to find a partition placement function $\mathcal{M}$, that maps the active partitions to VMs in each superstep, such that the actual cost $\Gamma$ to execute the application on the graph is minimized while also minimizing the increase in the actual makespan $T$ above the minimum makespan $T_{Min}$, i.e., \emph{find $\mathcal{M}$ that minimizes $\Gamma$ and $(T-T_{Min})$.}

% if  partitions that will be active in each  that should execute on this graph, the algorithm to be executed on the graph. For each superstep of the algorithm, determine the best possible placement scheme for active partitions in the superstep, in such a way that cost for the VMs to be used is minimised.\\

% \ysnote{the earlier para is not relevant since we're not talking about load imbalances due to partitioning, but imbalances due to non stationary algos. Also, we do not reduce the makespan as part of this paper, only reduce cost.}

% \ysnote{Give a formal definition of the problem. What are you given (graph, metagraph, partitions, number of supersteps, time a partition is active in each superstep)? What should you return (number of VMs at each superstep, and mapping of partitions to VMs)? What should you optimize (reduce cost, avoid increasing makespan)?}

%      In this paper, our main focus is given the analytical tools such as metagraph, how we can make effective use of such tools to model the behaviour of algorithm across the supersteps in terms of running time of different partitions in subgraph-centric frmework like GoFFish. This model will allow us to place these partitions on virtual machines in cloud. Our goal would be without disturbing the BSP proccesing model, place these partitions on VM's such that           the VM core minutes required for the execution of the algorithm is minimized  with secondary objective : Makespan of the algorithm is minimized.
As we shall discuss next, this can be modeled as an optimization problem, similar to bin packing, with the primary goal of reducing the cost of executing the application on elastic VMs through intelligent partition placement, and the secondary goal of reducing the increase in makespan above the theoretical minimum makespan.

% \ysnote{For IPDPS: what are stationary and non stationary algos? Give e.g. of each}
% \ysnote{Classify the different types on non-stationary. Give fig. Give examples, BFS, SSSP, BC, others.}

\section{Partition Placement Strategies }
\label{sec:elastic}

\subsection{Default Strategy}
The default ``flat'' partition placement strategy used in GoFFish is to allocate as many cores (VMs) as the number of partitions, with each 1-core VM exclusively operating on a single partition for all supersteps of the application~\cite{goffish-arxiv}. This placement function is given as: \[ \mathcal{M} : P_{i} \times s \rightarrow \upsilon_i \]
where each partition $P_i$ is placed in its own VM $\upsilon_i$. The advantage of this is that a partition is processed in as fast as manner as possible on its own VM, given the partition-centric programming model, and its makespan matches $T_{Min}$. It is also trivial to solve and implement this strategy in practice, taking $\mathcal{O}(n)$ in time complexity for $n$ partitions. 

The actual billing cost for this strategy corresponds to the $\Gamma_{Max}$, and since all $n$ VMs are kept active for the entire duration of the application, the cost can be simplified as: \[ \Gamma = n \cdot \ceil[\Big]{\frac{T_{Min}}{\delta}} \cdot \gamma \]

% ere, all partitions are assigned to a static number of VMs, and this mapping is retained for all supersteps.

% Default over-allocates VMs, and hence takes up more resources.

% Since partition is the granularity of computation, and each partition has its own VM, this is guaranteed to give the smallest makespan.

% \ysnote{For all algos, have these 4 parts}
% \paragraph{Approach and benefits}

% \paragraph{Algorithm Pseudo-code}

% \paragraph{Algorithm Description}

% \paragraph{VM Cost Function}

% \paragraph{Computational Complexity}

\subsection{Optimal and First Fit Decreasing Heuristics (OPT, FFD)}
%\paragraph{Approach and benefits}
We reduce the given problem to a linear programming problem that gives the optimal number of VMs and the mapping of partitions to those VMs for each superstep, while \emph{guaranteeing} that the makespan does not increase beyond the $T_{Min}$. We set the necessary condition to retain the makespan at $T_{Min}$ by requiring that each superstep takes only $\tau_{Max}^s$, which is the most time taken by any single partition in a superstep if it has an exclusive VM. This is given as:
\[ \max_{j=1}^{l}{ \big( \sum_{i=1}^n \widehat{\mathcal{M}}(i,s,j) \cdot \tau_i^s \big)} ~~ = ~~ \tau_{Max}^s,~~~\forall 1 \le s \le m \]

If we treat each VM as a bin that has a time capacity of $\tau_{Max}^s$ in superstep $s$, we need to find a mapping $\mathcal{M}$ for all partitions (given by their timing $\tau_i^s$) onto the \emph{smallest number of bins in each superstep}. We can use linear programming to solve this problem.

Two simplifying assumptions are made here to assure optimality. One is that there is no cost to reassign partitions to VM between supersteps. This is unlikely in practice as data movement of partitions between VM does incur time cost, which affects both the makespan and the billing cost. But it is useful to consider this optimal solution as a baseline for comparison. 
% this is a Linear programming problem. Solving this linear problem, would give us the minimum number of VMs required for each superstep. Although, solving LP is a costly operation, it would be useful to compare it with other strategies. 

% \paragraph{Algorithm Pseudo-code}
% Linear Programming formulation of problem :\\

% minimize cost $C$ = $\sum_{i=1}^n v_i * c_i$\\

% Here $c_i$ is cost of running VM $i$\\

% subject to 	$C \geq 1$\\

% $\sum_{j=1}^n P_j^s$  $ x_{ij} \leq SS_t * v_i,   \forall j \in \{1,\ldots,n\}$\\

% This condition indicates that, for each active VM the sum of partition time should not exceed capacity($SS_t$) of VM. $P_j^s$ is execution time of partition $j$ in superstep $s$\\

% $\sum_{i=1}^n x_{ij} = 1, 	 \forall j \in \{1,\ldots,n\}$    such that $P_j^s \ge 0 $\\

% This condition indicates that no active partition should be left unmapped.\\

% $ 	v_i \in \{0,1\}, 	\forall i \in \{1,\ldots,n\}$ \\

% This variable will be 1 if VM $i$ is active or 0 otherwise.\\

% $	 	x_{ij} \in \{0,1\}, 	\forall i \in \{1,\ldots,n\} \, \forall j \in \{1,\ldots,n\}$\\

% This variable is 1 when item $j$ is mapped to VM $i$.\\

% %%%%
The linear programming problem can be solved using standard techniques~\cite{lp-solve}, but it can be computationally costly to get the optimal solution. We term this strategy as Optimal (OPT). In addition, we use the heuristic First Fit Decreasing (FFD) algorithm to approximately solve this optimization problem with a lower complexity. 

The pseudocode for FFD is given in Alg.~\ref{alg:ffd}, and it follows a greedy approach. At each superstep $s$, we start with no existing VMs and the partitions $P_i$ are sorted in decreasing order of their execution times for that superstep, $\tau_i^s$. In this sorted order, we test if each partition's execution time can fit in an existing VM. If so, the partition is mapped to that VM for this superstep, and the VM's capacity decreased by that partition's execution time. If no VMs can hold this partition, we create a new VM with capacity $\tau_{Max}^s$ and map the partition to this new VM, and decrement its capacity.

%%%%
% In FFD, we allow each active partition to migrate to any VM in a superstep, with zero cost associated with data movement of the partition between VMs. We define the capacity of VM as the time required for the current superstep in default case.
%
%%%%
\textbf{Time Complexity.} OPT guarantees that the makespan does not increase more than $T_{Min}$ while minimizing the number of active VMs per superstep, which is expected to reduce the billing cost. The tight theoretical bound for FFD relative to OPT on the number of active VMs is $(11/9 \cdot OPT + 6/9)$~\cite{dosa2007tight}, but is faster to calculate.
Sorting the partitions requires  $\mathcal{O}(n\log{}n)$ time and mapping each to a VM requires a linear scan of all VMs. For the $i^{th}$ partition, a maximum of $i$ VMs can exist. So VM mapping takes:
\[ \sum_{i=1}^{n} \log{}i = \mathcal{O} (\sum_{i=1}^{n} \log{i} ) = \mathcal{O} (  \log{n!} ) = \mathcal{O} (n \log{n})\]
and each superstep takes $\mathcal{O} (n \log{} n)$. The total complexity of FFD for $m$ supersteps is $\mathcal{O} (m \times n \log{n})$\\

%%%%
% Note that this algorithm is a theoratical. It allows to migrate each active partition to any VM, which has a data movement cost associated with it. Since we have not taken into account, the data movement cost, it is not a good candidate for practical comparison. For evaluation and analysis, next we consider a variant where a penalty is paid for data movement. 

% \ysnote{The goal of OPT is to reduce the number of Active VM with the constraint that makespan should be equal to that of default exectution.}
% \drnote{Instead of sorting the VMs for each partition to be  mapped, just find the VM with maximum capacity}
%%%%

%\paragraph{Algorithm Pseudo-code}

\begin{algorithm}
\caption{First Fit Decreasing algorithm}\label{alg:ffd}
\begin{algorithmic}[1]
\Procedure{FirstFitDecreasing}{$P,\mathcal{A},n,m$}
\Comment{$P$ is the set of $n$ partitions. $\mathcal{A}$ is the time function that gives $\tau$ values. $m$ is the number of supersteps.}
\For{$s \le m$} \Comment{iterate over supersteps}
%\State $C_j^s \gets \tau_{Max}^s$ \Comment Capacity of VM $j$ in $s$
\State $v[~] \gets \varnothing$ ; $l=0$ \Comment{init list of VM capacities}
\State $p[~] \gets \textsc{SortDescending}(P)$ \Comment{Sort by $\tau_i^s$}
\For{$i \le n$ } \Comment{iterate over each partition}\label{alg:ffd:eachpart}
\State assigned $\gets false$
\For{\textbf{each} $j \le l$ \textbf{and} assigned $= false$ }\label{alg:ffd:eachvm}

\Comment{does VM $j$ have capacity for partition $i$?}
\If { $v[j] \ge \tau_{p[i]}^s $ }
\State $\mathcal{M}(p[i],s) \gets j$ \Comment {map $P_i$ to $v_j$}
\State $v[j] = v[j] - \tau_{p[i]}^s$ \Comment{$\downarrow$ capacity}
\State assigned $\gets true$
\EndIf
\EndFor
\If {assigned $= false$}
\State $v[++l] \gets \tau_{Max}^s$ \Comment{create new VM}\label{alg:ffd:taumax}
\State $\mathcal{M}(p[i],s) \gets l$ \Comment{do mapping}
\State $v[l] = v[l] - \tau_{p[i]}^s$ \Comment{reduce capacity}
\EndIf
\EndFor \Comment{mapping done for superstep $s$}
\EndFor \Comment{mapping done for all supersteps}
\State \Return $\mathcal{M}$
\EndProcedure
\end{algorithmic}
\end{algorithm}

%\paragraph{Algorithm Description}

%\paragraph{VM Cost Function}
\label{sec:activation}\textbf{Activation Strategy.} Given the mapping function from the placement strategy, we propose a VM activation strategy to minimize the actual billing cost. This decides whether each VM, at the end of a superstep, can be left running or terminated for the next superstep. This is important since billing rounds up to the nearest $\delta$, and stopping a VM for less than $\delta$ time and restarting it is costlier than retaining that VM idly for that duration, i.e., if $l$ VMs are used in a superstep $s$, $l-1$ in superstep $s+1$, and again $l$ VMs in $s+2$. If the duration of superstep $s+1 \le \delta$, it is cheaper to retain $l$ VMs for all 3 supersteps.

In our strategy, for each VM we do such a test, to see if the time remaining in a VM before the next $\delta$ increment is less than the time taken by the next superstep. If so, we retain that VM and reuse it for that superstep rather than create a new VM. Otherwise, if keeping that VM will cause it to go past the $\delta$ boundary, we terminate it.

%Consider the following example. A graph algorithm is running on cloud system using $n$ VMs and current superstep is $i$. Now in next superstep, suppose the algorithm mapped all partitions on only $n-1$ VMs. In this case, for the one remaining VM, which has no partitions mapped on it, we can either terminate it or keep it idle. Suppose that if we terminate the VM at the end of $i^th$ superstep, we have to pay for $m$ core minutes.  Our cost function, takes into consideration, the time for next superstep $i+1$. If the time for $i+1$ superstep added to the current VM execution time sums upto $m$ core minutes, then we will keep the VM idle otherwise terminate it. 

% \begin{algorithm}
% \caption{Cost Function}\label{costFunction}
% \begin{algorithmic}[1]
% \Procedure{Cost function}{VMTime $VM_t$, SuperstepTime $SS_t$}\\

%   $presentCoreMin=\ceil[\big]{VM_t}$
  
% \If{$presentCoreMin == \ceil[\big]{VM_t + SS_t}$}
%    \State Keep the VM idle
% \Else
%     \State Terminate the VM
% \EndIf   
    
% \EndProcedure
% \end{algorithmic}
% \end{algorithm}

% \paragraph{Computational Complexity}

\textbf{Data Movement Cost.} As a variation of OPT, we also evaluate these mapping if the data movement cost for moving a partition from one VM to another is considered. Called OPT-DM, this is more realistic when implementing OPT (or FFD) on the Cloud without special means to rapidly move or mount partitions between VMs. Here, the placement algorithm itself remains the same, but when calculating the billing cost, we include the time for data movement that causes the VM's billing to increase. 

%\subsection{OPT and FFD with Data Movement Cost}
%
% \paragraph{Approach and benefits}
% This algorithm is very similar to the FFD algorithm, except that, it assigns a data movement cost with each mapping of partition to a VM, for every superstep. This algorithm is practical implementation of FFD algorithm and thus, is useful for practical comparison with other approaches.
%
%\paragraph{Data Movement Cost}
Specifically, we assume a shared persistent storage (like AWS S3 or Azure BLOB store) where partitions are moved to from VMs at the end of each superstep, and then before the next superstep starts, they are copied to the set of VMs that they are mapped to in that superstep. So each VM pays the time cost for moving data in and out of it at the start and end of a superstep, and is added to the billing cost. % Also at the end of the superstep, the partitions are dumped to the S3 storage. Question arises, why we are not directly moving it to new VM where it would get mapped? Because the mapping is done at run time, and the VM to which the partition would get mapped to is not known ahead of time. So it is not possible to initiate the migration ahead of the start of superstep.

\subsection{Max Fit packing with Pinning (MF/P)}
In this strategy, we avoid the data movement costs of OPT-DM by ``pinning'' a partition to a particular VM, and not changing the mapping after that. In other words, for a partition $P_i$ whose $\tau_i^s > 0$:
\[\mathcal{M}:P_i \times s \rightarrow \upsilon_j \implies \mathcal{M}:P_i \times s' \rightarrow \upsilon_j,~~\forall s' > s \] 
and hence, $\widehat{\mathcal{M}}(i,s,j)=1 \implies \widehat{\mathcal{M}}(i,s',j)=1,~~\forall s' > s $

Here, we use a strategy similar to FFD to greedily place an unpinned partition onto an existing VM with the \emph{maximum} available capacity in a superstep, if possible, and if not possible, start a new VM. Once pinned, the partition remains in that VM for the rest of the application. As a result, there are no data transfer costs, which additionally makes this simpler to implement.%  since data does not have to be checkpointed and restored during data transfer.
We term this as \emph{Max Fit with Pinning (MF/P)}. 

In FFD, the capacity of a VM in a superstep $s$ was $\tau_{Max}^s$. However, some of the VMs at the start of a superstep may already have partitions pinned on them, some or all of which may be active in this superstep. As a result, the makespan of this superstep depends both on the largest (unpinned) partition time, as well as the VM holding the largest cumulative pinned partition times. 

Let $\lambda_j^s$ be the load on a VM $j$ in superstep $s$, defined as: $\lambda_j^s = \sum_{i=1}^n \widehat{\mathcal{M}}(i,s,j) \cdot \tau_i^s$, i.e., the cumulative time of all partitions mapped to that VM. We redefine $\tau_{Max}^s$ for MF/P at the start of superstep $s$ as:
\[\tau_{Max}^s = \max\Big(~~ \max_{i=1}^n\big( \tau_i^s \big),~~ \max_{j=1}^{l}{\big( \lambda_j^s \big) }~~ \Big)\]
Here, the first term within the outer \emph{max} function gives the time taken by the largest partition in this superstep, while the second term gives the largest of the total times taken by all pinned partitions in a VM.

We skip the pseudocode for MF/P for brevity. Its key distinctions from FFD are that the partitions do not migrate between VM across supersteps, the initial capacity of a VM on a superstep is based on the largest partition in that superstep as well as the pinned partitions, and we pick the VM with the largest capacity for partition placement, rather than the first VM that has adequate capacity. This results in the following changes to Alg.~\ref{alg:ffd}. In Line~\ref{alg:ffd:eachpart}, we only iterate through partitions that are not already pinned in a previous superstep. Those that are pinned retain their mapping. In Line~\ref{alg:ffd:eachvm}, rather than iterate through each VM's capacity, we only test the VM with the largest available capacity. And lastly, in Line~\ref{alg:ffd:taumax}, we compute the value of $\tau_{Max}^s$ based on the updated function given above.

Note that this strategy does not guarantee a makespan that matches $T_{Min}$, as is obvious from the higher value of $\tau_{Max}^s$. We retain the same VM activation strategy as used in OPT to decide whether to keep a VM active or terminate it at the end of every superstep.
We retain the same VM activation strategy as used in OPT to decide whether to keep a VM active or terminate it at the end of every superstep.

\textbf{Computational Complexity.}
The MF/P strategy maps each partition exactly once and keeps this mapping throught the runtime of the application. For mapping a partition, it finds the VM with the maximum capacity. If the partition fits in that VM, it maps the partition to that VM and otherwise spins a new VM.  Finding the VM with the maximum capacity requires linear time. So the asymptotic time complexity of this algorithm across all supersteps is $\mathcal{O} (n^2)$.

%Once the mapping is determined for all partitions, a constant time will spend in each for a superstep. If the number of supersteps is high, this time will make effect. So the total time complexity is $\mathcal{O} (n^2) + \mathcal{O} ( \delta )$

% %%%%
% \paragraph{Data Movement Cost}
%  In case of FFP, once the mapping for a partition is determined, same mapping is followed in subsequent supersteps. So, unlike the FFD algorithm, FFP does not require the partitions to be read and write back to the S3 storage in every superstep. So data movement cost is not applicable here.

\subsection{First Fit Lookahead, with Pinning (LA/P)}
%\ysnote{RAVI: Check if the ranking can be done purely based on the incremental pinned load in the next superstep}
One of the potential downsides of MF/P is that it decides to pinning a partition to a VM based only on the time taken by the partition in the current superstep. Since partitions once pinned do not migrate, we may end up with a placement that may be well-suited in the current superstep but lead to under-performance in future supersteps. This can result in several VM with pinned active partitions that are unbalanced in a superstep, which can increase the makespan. Since the \emph{a priori} prediction model can provide partition timings for all supersteps, we can leverage this for a more global planning across supersteps. To keep the problem tractable, we propose a variation of MF/P where the partition information for the current superstep and the next superstep are used to decide placement. Once decided, we continue to pin a partition to a VM. We term this strategy as \emph{Lookahead with Pinning (LA/P)}.

The intuition here is that we first map unpinned partitions in a superstep, going from partitions the largest to the smallest execution times. Then, when considering VMs to map them to, we prefer VMs that have a higher capacity in the \emph{next} superstep, rather than consider the first VM with adequate capacity (FFD) or the VM with the largest capacity (MF/P), in the current superstep.
% \paragraph{Approach and benefits}
% Here, we greedily place a partition on the first VM that it is active on, but also consider how loaded the VM is in the next superstep when deciding the binning. We do not change the placement.

% This can potentially deal with cases where packing partitions in a single VM causes over-allocation in the next superstep due to partitions active in this superstep also being active in the next superstep. This lookahead can reduce the chance of makespan increasing as a result of this.

% \paragraph{Algorithm Description}
We use two \emph{rank} values to decide this placement: % Before we introduce the actual algorithm, we define two new terms 'Current Rank' and 'Forward Rank'. Both terms specify an integer value for each active partition and active VM respectively.

%%%%
\textbf{Current Rank} % ($C_i^s$)
for each \emph{active} partition $P_i$ (i.e. $\tau_i^s >0$) in superstep $s$ is the index of that partition when all the active partitions are sorted in descending order of the execution times, $\tau_i^s$.
% Sort the active partitions, based on the execution time in superstep $i$ in decreasing order. The Current Rank of partition $P_k$, for $i^th$ superstep, is it's index in the sorted list.  

%%%%
\textbf{Forward Rank} % ($F_i^s$) 
for each \emph{active} VM $v_j$ in superstep $s$ (i.e., $v_j$ has some active partition in $s$ pinned to it) is the index of that VM when all the VMs are sorted in ascending order of their load in the \emph{next} superstep, $\lambda_j^{s+1}$.

% Perform mapping as per \emph{partition placement function} $\mathcal{M}$ and calculate Load for each active $V_k$, in superstep $i+1$.

%   \[ L^k = \big( \sum_{j=1}^{m} P_j^{s+1} \big) \]
  
%   where, $ P_j^{s+1} = \{P_j ~|~ \exists V_k : (P_i \times s+1 \rightarrow V_k ) \}]$

%  Sort the VMs, in decreasing order, based on the Load value. The Forward Rank of the VM is given by the index of VM in the sorted list.

Here again, we describe LA/P in terms of its distinctions from FFD's Alg.~\ref{alg:ffd}. In Line~\ref{alg:ffd:eachpart}, we first sort the partitions based on their current rank and only iterate unpinned partitions. Those that are pinned retain their mapping. In Line~\ref{alg:ffd:eachvm}, we sort the VMs based on their forward rank and then iterate through them. The VM mapping, however, is done only if the VM has adequate capacity in the current superstep. The forward rank is also recalculated after each new mapping in this superstep. And lastly, in Line~\ref{alg:ffd:taumax}, similar to MF/P, we compute the value of $\tau_{Max}^s$ based on the updated function that considers pinning.

In LA/P too, we do not guarantee that the makespan does not grow beyond $T_{Min}$, and we use the same VM activation strategy as FFD. 
%The cost function defined in \ref{costFunction} is applied here to minimize the cost.

\textbf{Computational Complexity.}
For each superstep, the algorithm sorts $n$ partitions based on their execution time to calculate their current rank, taking $\mathcal{O}(n\log{n})$ time. For each of the $n$ partitions to be mapped, the VMs are sorted based on their forward rank, that takes $\mathcal{O}(n \log{n})$ time. This gives a total complexity of $\mathcal{O}(n\log{n} + n^2\log{n}))$, which is dominated by the latter term. %  = \mathcal{O} (n \sum_{i=1}^{n}  \log{}i ) = \mathcal{O} (  \log{}n! ) = \mathcal{O} (n^2 \log{} n)$$

\section{Evaluation}% \note{2.5pgs}}
\label{sec:evaluation}
We evaluate the different placement strategies for performing SSSP (BFS) over undirected graphs on elastic VMs. We initially run the graph algorithm over the different graphs on a commodity cluster to get their partition timings ($\mathcal{A}$), and use this as input for simulating the application execution using the strategies. %We discuss the datasets, setup and evaluation metrics next.

\subsection{Setup and Datasets}
We use $4$ graphs that are partitioned into 8 or 40 partitions, as noted: LIVJ/8P~\footnote{http://snap.stanford.edu/data/soc-LiveJournal1.html}, USRN/8P~\footnote{http://www.dis.uniroma1.it/challenge9/download.shtml}, and ORKT/40P~\footnote{http://snap.stanford.edu/data/com-Orkut.html}.
\begin{table}[h]
\centering
\caption{Datasets used}
%\footnotesize
\label{tbl:datasets}
\begin{tabular}{lrrr}
\toprule
\textbf{Graph (Name/Part.s)}  & \textbf{|V|} & \textbf{|E|} & \textbf{Dia.} \\
\hline
\hline
LiveJournal (LIVJ/8P) & $4.847$M      & $68.993$M  & $16$             \\ %\midrule
USA Road (USRN/8P)       & $23.947$M     & $58.333$M  & $6,262$           \\ %\midrule
Orkut (ORKT/40P)                  & $3.072$M        & $234.370$M   & $9$   \\ \bottomrule
\end{tabular}
\end{table}
%
%%%%
For the default strategy, we run the subgraph-centric SSSP/BFS application using GoFFish on a 24-node commodity cluster connected by Gigabit Ethernet, and each partition is allocated one AMD~3380 2.3~GHz core and 4~GB RAM, and shares a 256~GB SSD on that node. We use CentOS 7 and JDK v7. % For the LIVJ/8P and USRN/8P graphs, 2 physical machines were used with each machine holding 4 partitions each. Whereas for ORKT/40P, we used 5 machines such that each machine has 8 partitions each.
The graphs are partitioned using METIS with a default load factor of 1.03 for vertex-balanced partitioning.

%%%%
We use GoFFish's logging framework to get the compute time for each subgraph in a superstep, and calculate the time for a partition as the sum all its subgraph times in that superstep. % accumulate the  in each superstep. \drnote{We have ignored time for messaging and synchronization overhead}We then calculated the compute time, required for each partition, by taking the sum of compute times for all subgraphs, belonging to that partition. 
This is used as the partition execution time ($\mathcal{A}$) passed to the strategies. %, required for partition in corresponding supersteps, in algorithms described in section $5$.

%%%%
The placement strategies are scripted in Python v2.7, and take as input $\mathcal{A}:P_i \times s \rightarrow \tau_i^s$, along with the number of partitions $n$ and the supersteps $m$. % These values of $\tau_i^s $ are the execution time for each partition, in each superstep.
The output generated by the algorithms is the partition placement function $\mathcal{M}:P_i \times s \rightarrow \upsilon_j$. From this mapping, we calculate the actual makespan, billing cost, and other metrics described next. %information such as makespan, cost for the VMs, utilization etc.

% \ysnote{RAVI: describe the machines and framework used to generate inputs (i.e. $\mathcal{A}$ function), how the algos were scripted (Python?), how the simulation was done to get costing and time?}\drnote{done}

\subsection{Plots and Metrics}
\begin{figure*}[!t]
\centering%~%
  \subfloat[Makespan for LIVJ/8P]{
    \includegraphics[width=0.30\textwidth]{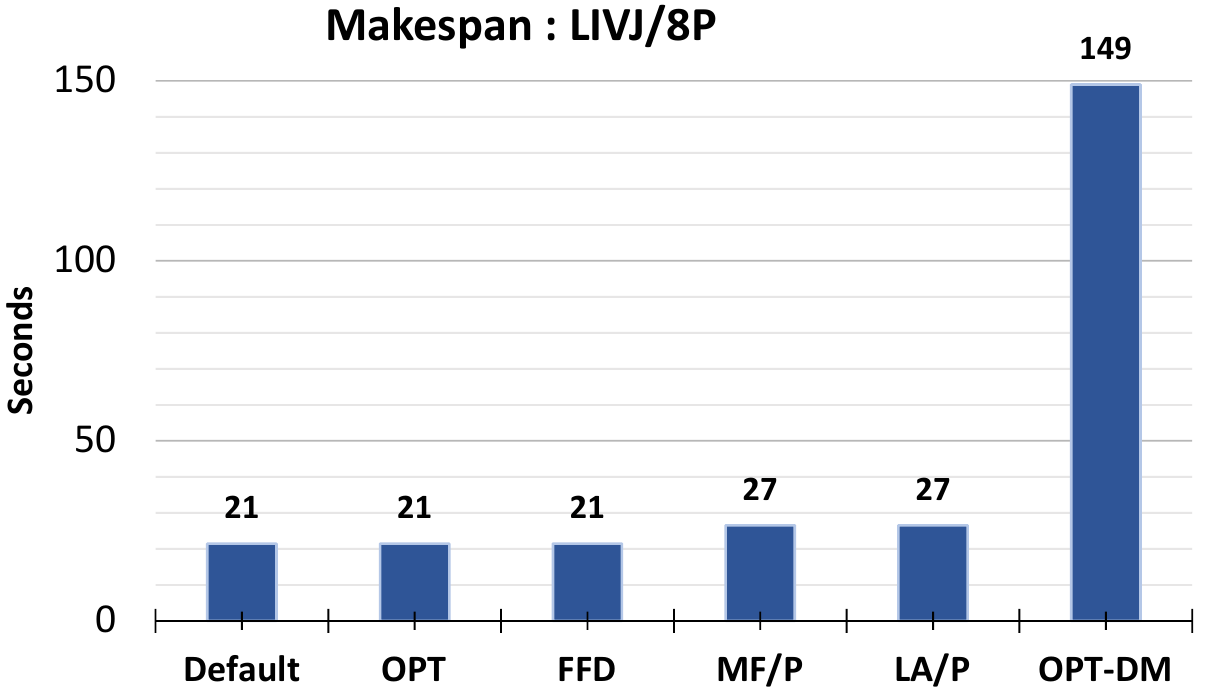}
    \label{fig:makespan:livj8}
  }
  \subfloat[Makespan for USRN/8P]{
    \includegraphics[width=0.30\textwidth]{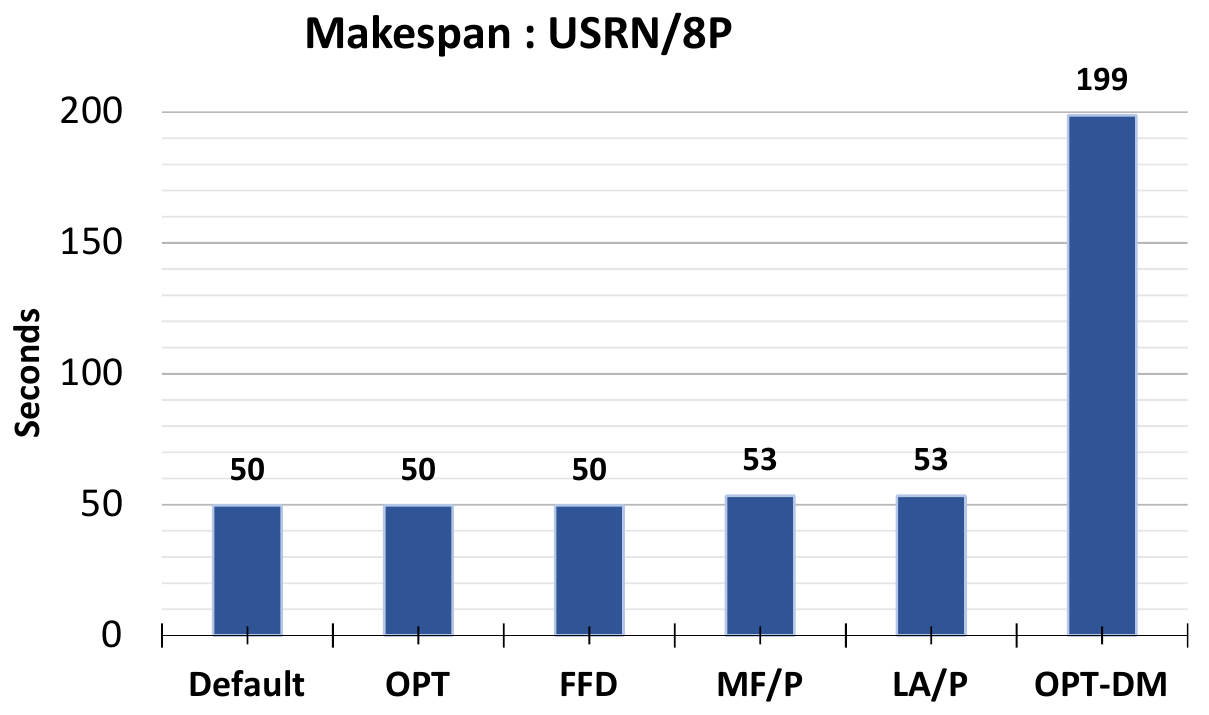}
    \label{fig:makespan:usrn8}
  }
  \subfloat[Makespan for ORKT/40P]{
    \includegraphics[width=0.30\textwidth]{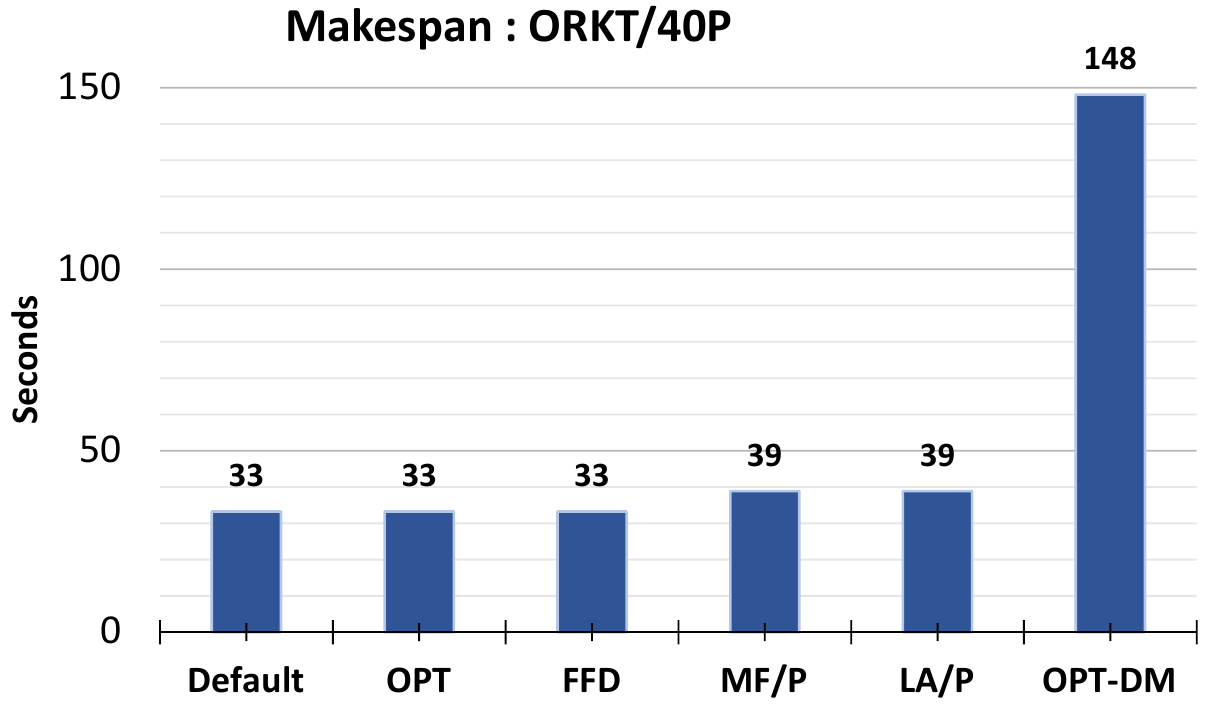}
    \label{fig:makespan:orkt40}
  }\\

  \subfloat[Core-Mins cost for LIVJ/8P]{
    \includegraphics[width=0.30\textwidth]{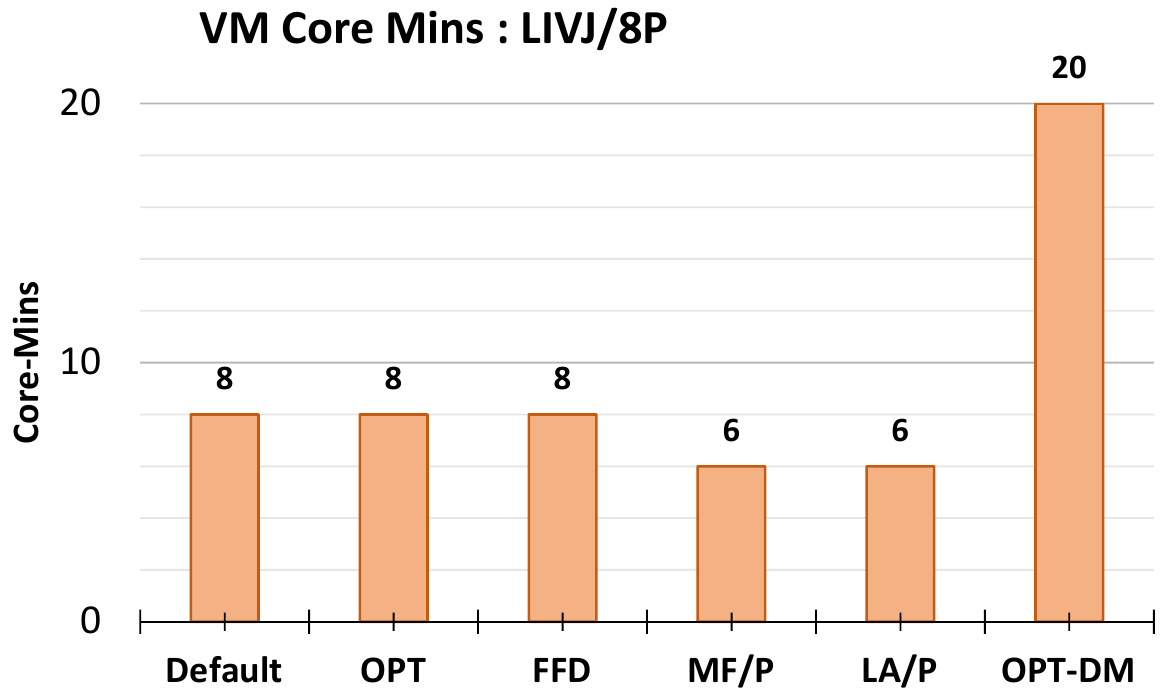}
    \label{fig:coremins:livj8}
  }
  \subfloat[Core-Mins cost for USRN/8P]{
    \includegraphics[width=0.30\textwidth]{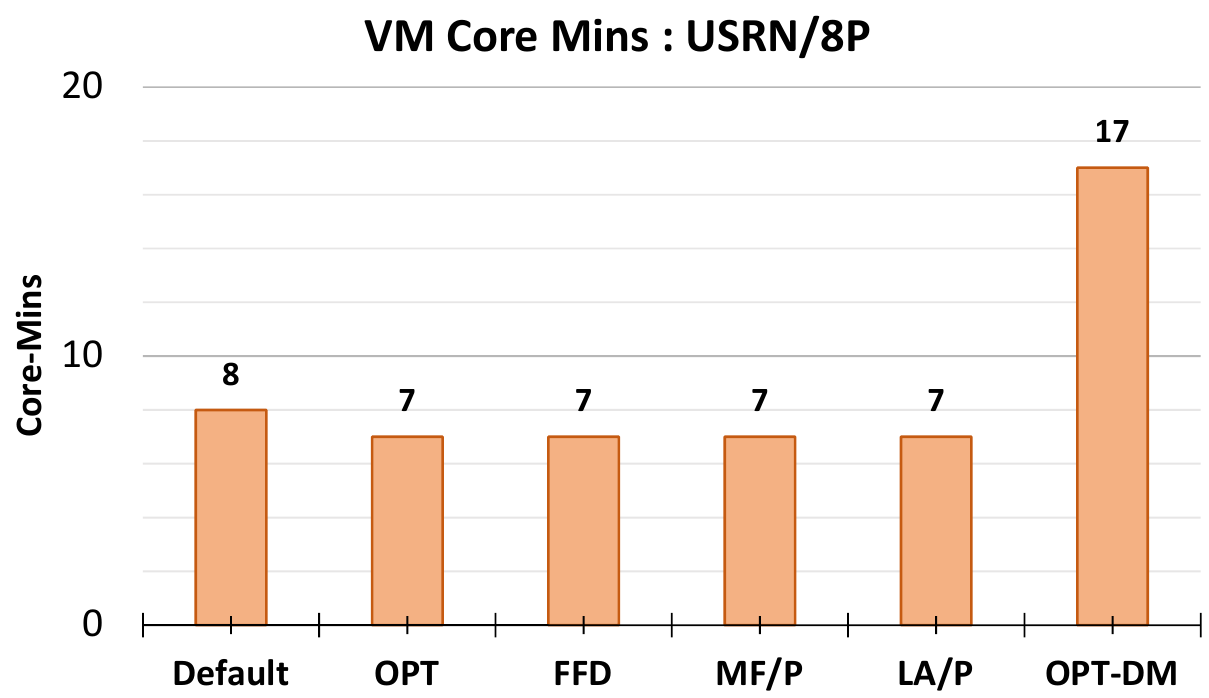}
    \label{fig:coremins:usrn8}
  }
  \subfloat[Core-Mins cost for ORKT/40P]{
    \includegraphics[width=0.30\textwidth]{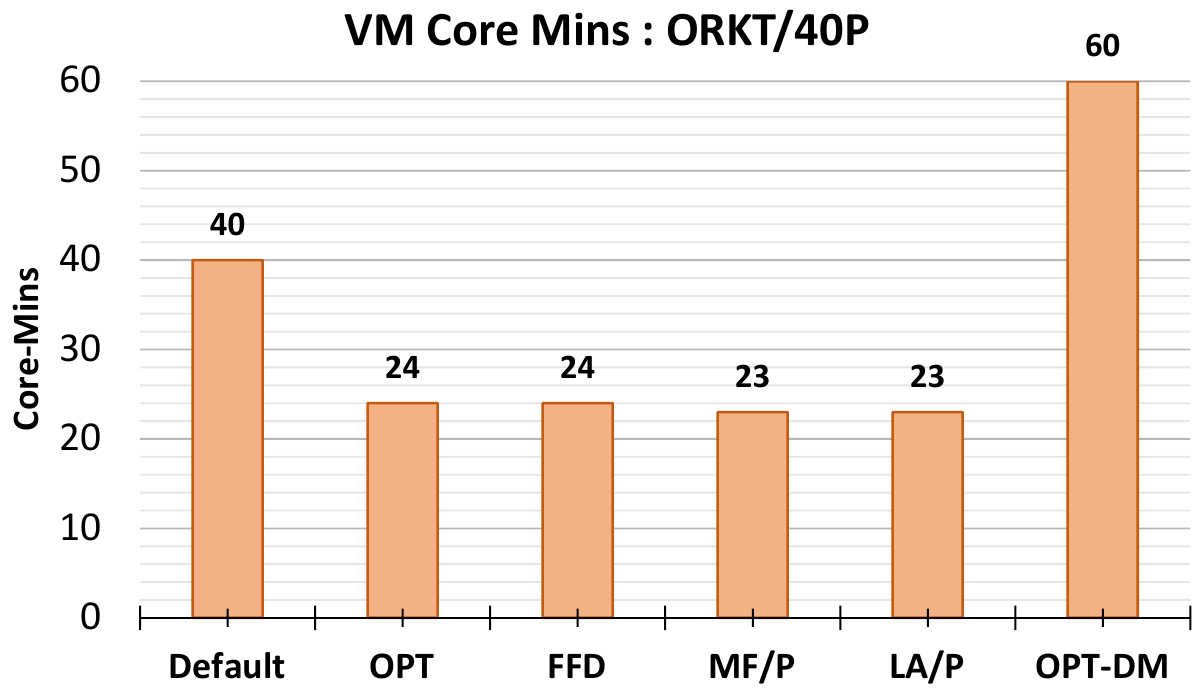}
    \label{fig:coremins:orkt40}
  }\\

  \subfloat[Under Utilization for LIVJ/8P]{
    \includegraphics[width=0.30\textwidth]{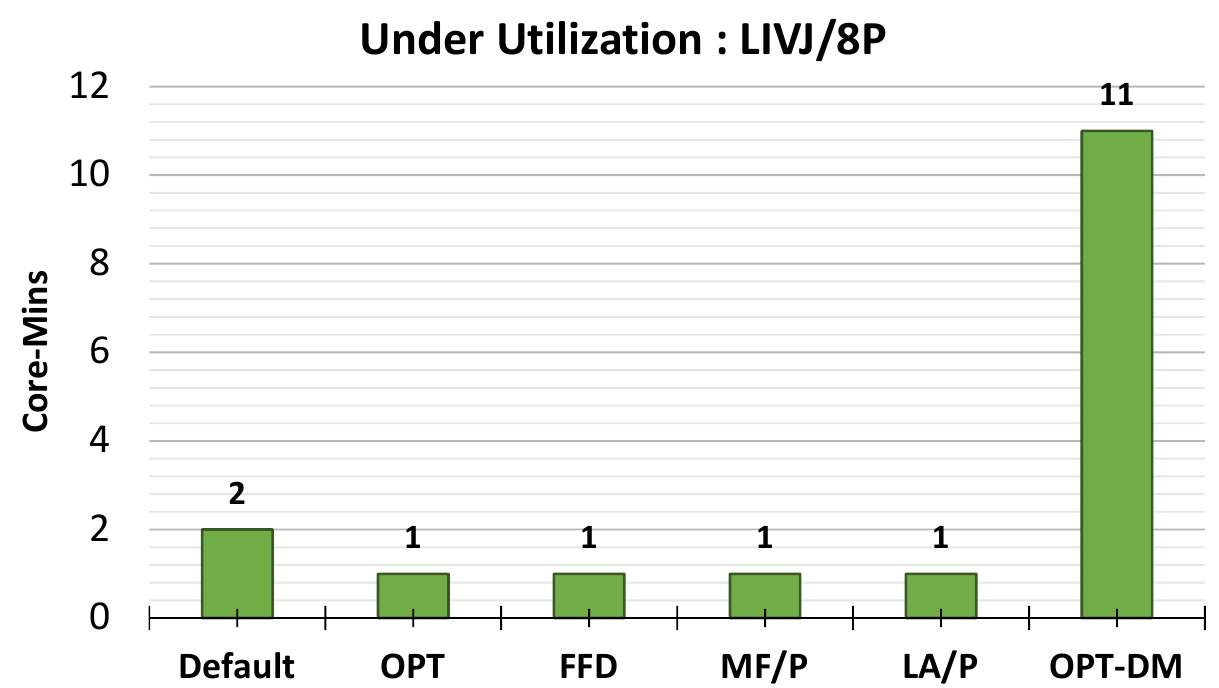}
    \label{fig:underutil:livj8}
  }
  \subfloat[Under Utilization for USRN/8P]{
    \includegraphics[width=0.30\textwidth]{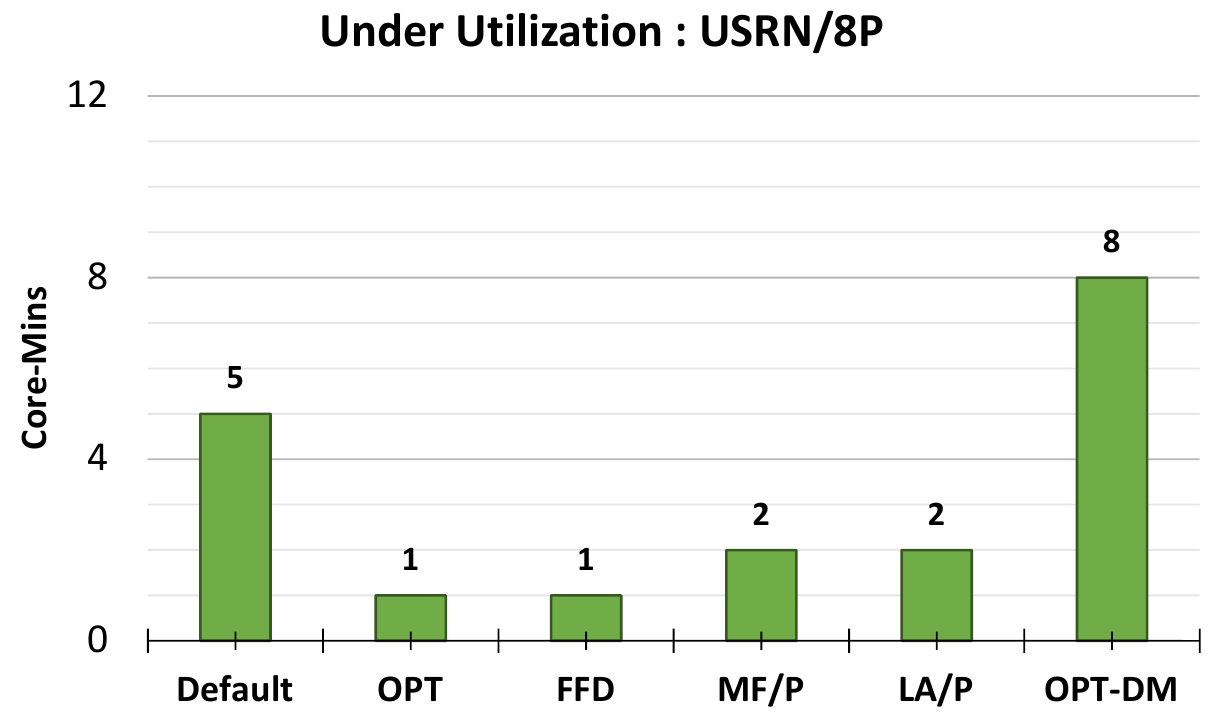}
    \label{fig:underutil:usrn8}
  }
  % \subfloat[Makespan for ORKT/16P]{
  %   \includegraphics[width=0.30\textwidth]{figures/makespan-orkt16.pdf}
  %   \label{fig:makespan:orkt16}
  % }
  % \subfloat[Core-Mins cost for ORKT/16P]{
  %   \includegraphics[width=0.30\textwidth]{figures/coremins-orkt16.pdf}
  %   \label{fig:coremins:orkt16}
  % }
  % \subfloat[Under Utilization for ORKT/16P]{
  %   \includegraphics[width=0.30\textwidth]{figures/underutil-orkt16.pdf}
  %   \label{fig:underutil:orkt16}
  % }\\
  \subfloat[Under Utilization for ORKT/40P]{
    \includegraphics[width=0.30\textwidth]{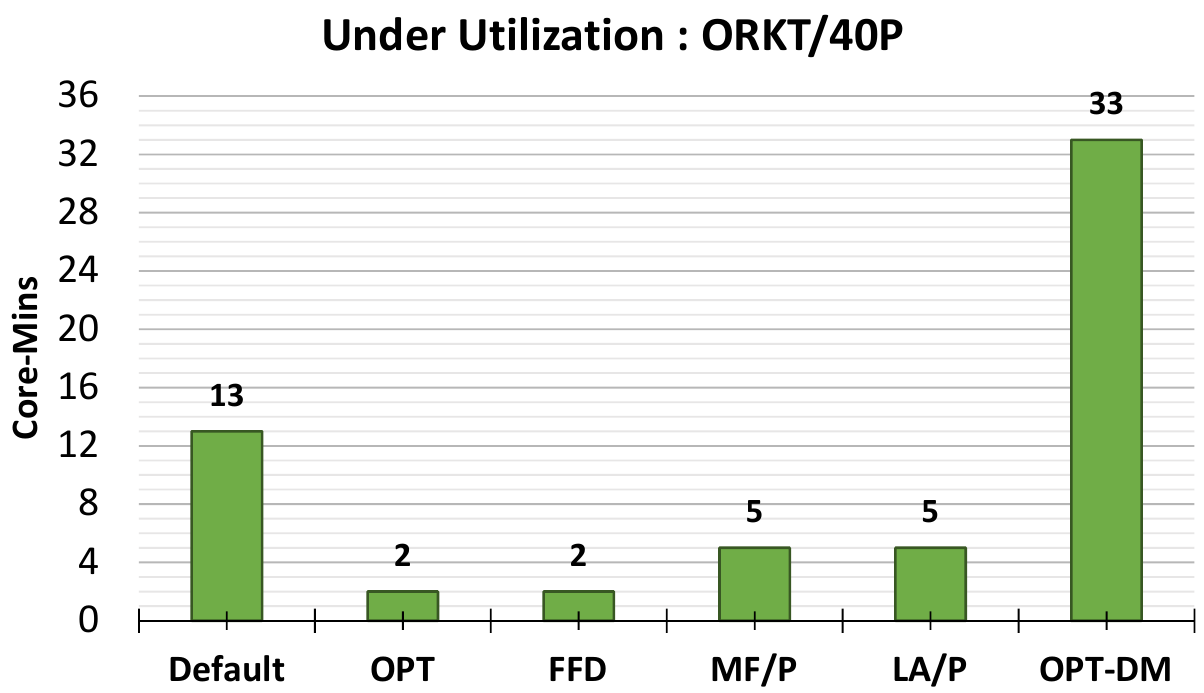}
    \label{fig:underutil:orkt40}
  }\\
  \subfloat[Core-Secs cost for LIVJ/8P]{
    \includegraphics[width=0.30\textwidth]{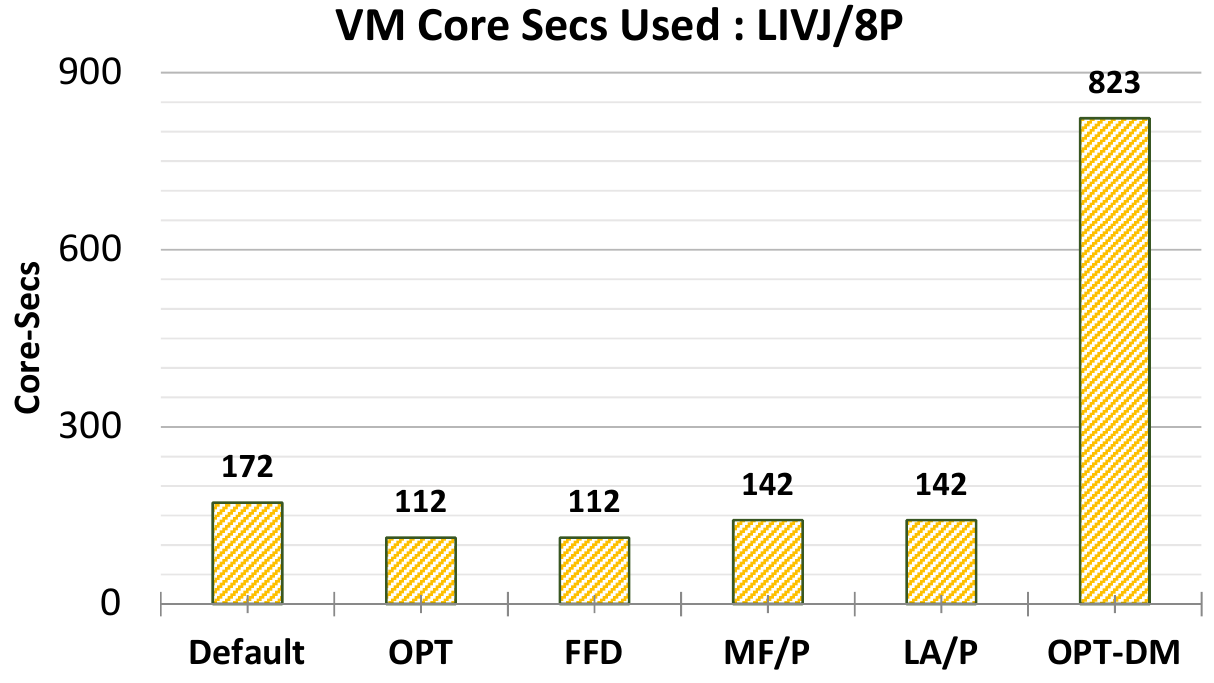}
    \label{fig:coresecs:livj8}
  }
  \subfloat[Core-Secs cost for USRN/8P]{
    \includegraphics[width=0.30\textwidth]{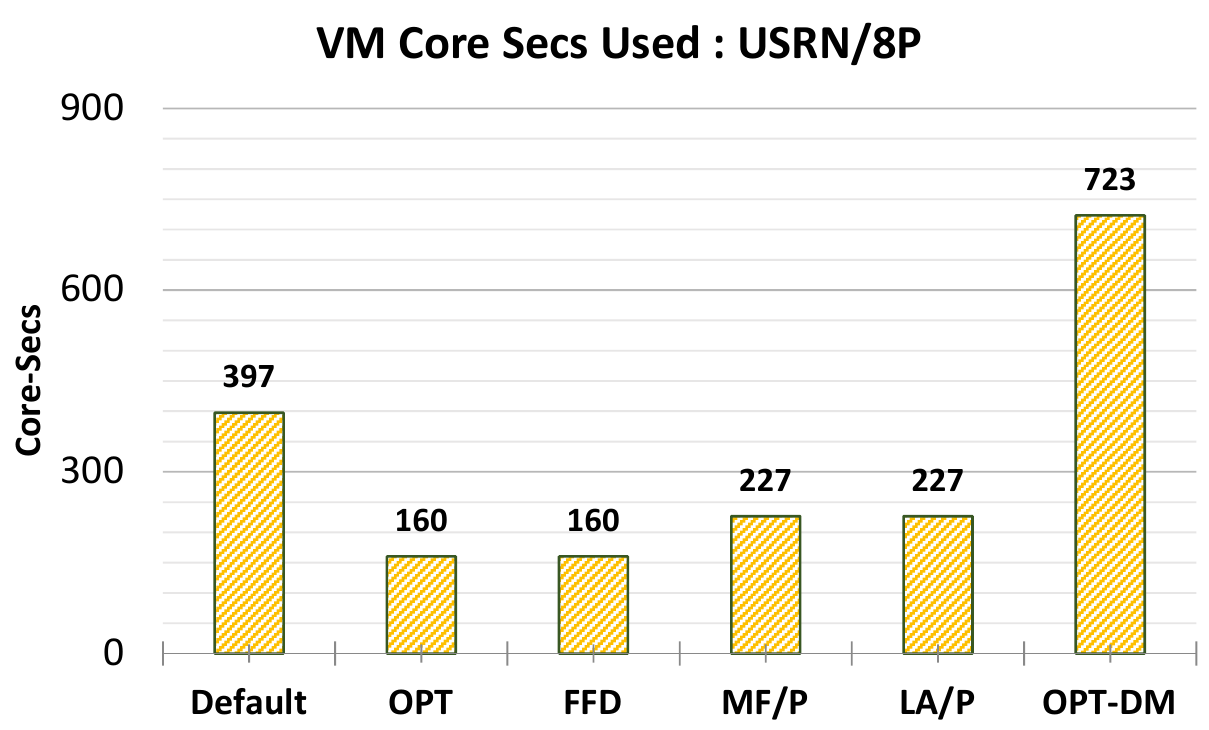}
    \label{fig:coresecs:usrn8}
  }
  \subfloat[Core-Secs cost for ORKT/40P]{
    \includegraphics[width=0.30\textwidth]{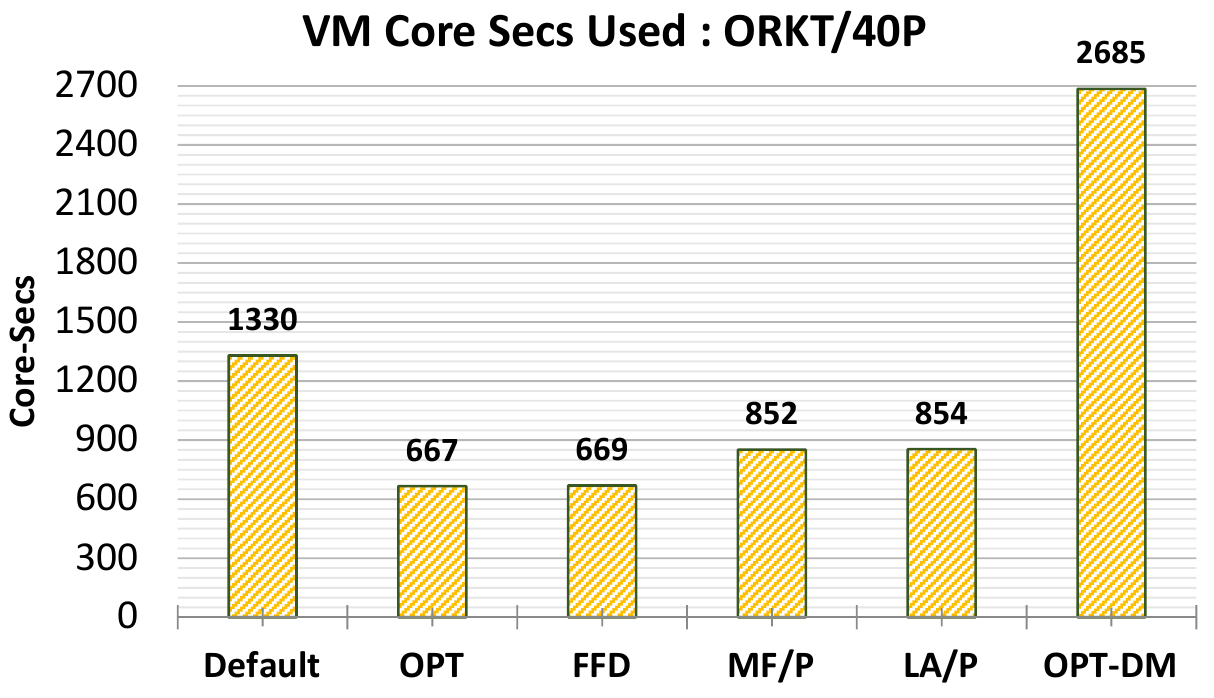}
    \label{fig:coresecs:orkt40}
  }

\caption{Performance metrics of the three graphs for different placement strategies.}
\label{fig:plots}
\end{figure*}

% \ysnote{RAVI: Describe the plots that we show and equations on how they were calculated.}\drnote{done}

%%%%
\textbf{Makespan:} Makespan is the total time taken by the graph application, as calculated for the mapping returned by each strategy. It is calculated as the sum over all supersteps of the time taken by the slowest VM in that superstep. This is plotted for each graph in Figs.~\ref{fig:makespan:livj8}--\ref{fig:makespan:orkt40}. % It gives the execution time required for the algorithm, when all VMs run in parallel as per BSP model of execution. $C^s$ denotes the maximum capacity of a VMs in superstep $s$.
% $$makespan= \sum_{i=1}^{n}  C^s$$
% where $n$ is the number of supersteps.

%%%%
\textbf{Cost in Core-Minutes:} While makespan gives the actual runtime of the application, the actual billing cost depends on how long each VM was active, and when they were turned on and off, as decided both by the mapping and the VM activation strategy. We use the mapping and activation information to calculate the total core-mins for which the VMs will be billed, using a 1-minute billing cycle, and rounding up the VM cost to the nearest minute each time it is turned off. This duplicates the actual billing logic used by IaaS Cloud providers like Azure. The actual monetary cost is just a multiplication of these core-mins by the per-minute rate, which depends on the data center, VM size, etc. %  At the end of the execution of the algorithm, we sum the running time of the VM over all superstep and ceil the obtained value to the nearest minutes value. 
This is plotted for each graph in Figs.~\ref{fig:coremins:livj8}--\ref{fig:coremins:orkt40}.

%%%%
\textbf{Under-Utilization:} In the BSP exection model used by component-centric frameworks, VMs remain active for the entire superstep even if some of them have completed processing their partitions. We capture the wasted time due to the slowest VM across all supersteps as under-utilization. It is given as the difference between the core-minutes for which VM were provisioned and the core-minutes for which they actually processed partitions, i.e.,
\[ \sum_{s=1}^m \big(\tau_{Max}^s \cdot \mid \Upsilon_s \mid \big) -  \sum_{s=1}^m \sum_{i=1}^n \big( \tau_{i}^s  \big)\] 
This is plotted for each graph in Figs.~\ref{fig:underutil:livj8}--\ref{fig:underutil:orkt40}.

% Let $\abs{ActiveVM}^s$ denotes the number of Active VMs in superstep $s$ and $C^s$ denotes the maximum capacity of a VMs in superstep $s$.  Let $m$ be the number of partitons such that $P_j^i$ denotes execution time of partition $P_j$ in superstep $s$.  The underutilization value is defined as 
% $$underUtilization= \sum_{i=1}^{n} \big( \abs{ActiveVM}^s \times C^s - \sum_{j=1}^{m} P_j^i \big) \ $$
% where $n$ is the number of supersteps.

%%%%
\textbf{Core-Seconds:} The core-mins cost uses a VM-minute granularity billing model provided by cloud providers. We attempt to highlight the relative benefits of the strategies if such costing granularity constraints were not present, and calculate the core-seconds for which VMs were provisioned by each strategy. This is given by,
\[ \sum_{s=1}^m \big(\tau_{Max}^s \cdot \mid \Upsilon_s \mid \big) \]
 % Let $\abs{ActiveVM}^s$ denotes the number of Active VMs in superstep $s$ and $C^s$ denotes the maximum capacity of a VMs in superstep $s$. The coreSec is defined as 
% $$coreSec= \sum_{i=1}^{n} \abs{ActiveVM}^s \times C^s$$
% where $n$ is the number of supersteps.
This is plotted for each graph in Figs.~\ref{fig:coresecs:livj8}--\ref{fig:coresecs:orkt40}.

\subsection{Analysis}

The time, cost and utilization values for OPT and FFD are identical in all cases. So the \emph{FFD algo is a good enough approximation} for OPT while only taking 1~sec to run the placement strategy for the largest graph ORKT/40P, compared to 13~secs taken to calculate using OPT. Hence, FFD can be chosen over OPT when performing online scheduling. %This holds for the DM variants, OPT-DM and FFD-DM, also.

Further, the makespan for OPT and FFD are the same as the default strategy in all cases and equals $T_{Min}$, the smallest possible makespan. Both these algorithms are successfully able to provide \emph{adequate VMs for the required computation on the active partitions} to allow them to complete without delay. Since they do not consider data movement costs in a superstep, the only time spent is on the computation of the active partitions by an exclusive VM. Hence, the secondary objective of not increasing the makespan above $T_{Min}$ is also achieved. However, in practice, the partitions will need to be moved between VMs across supersteps depending on the placement mapping generated, so OPT-DM is more plausible.

For \textbf{LIVJ/8P}, the makespan for MF/P is modestly higher than the default, at $27$~secs against $21$~secs, and is the same as LA/P (Fig.~\ref{fig:makespan:livj8}). 
We observed here that the VM with maximum capacity for MF/P and the VM with highest forward rank for LA/P turned out to be same, causing the mapping performed by both the algorithms to be identical.  
% \ysnote{RAVI: why are the makespans comparable here? Get the capacity of the VMs for each of these algos and see if they are similar, and hence cause the rank to be the same for both these algos.}

We also see that the core-mins for OPT, FFD, MF/P and LA/P are all comparable, taking $6$-$8$~core-mins (Fig.~\ref{fig:coremins:livj8}). % \ysnote{Why is it that FFD/P and LA/P have smaller core-mins than OPT and FFD? We may have discussed? In theory, should OPT always have a smaller core mins? Maybe the pinning decision causes fewer VMs to be active, hence causing the makespan to increase but also reducing the VM costs?}
This cost for MF/P and LA/P is smaller than the $8$~core-mins taken by the default strategy, saving $25\%$ in cost.

Interestingly, MF/P and LA/P cost lesser than OPT and FFD. This is because OPT and FFD guarantee that the makespan will not rise beyond $T_{Min}$, and as a result allocate adequate VMs to meet this goal. But MF/P and LA/P do not allow partition movement once they are pinned, and in this case, all partitions are pinned at the end of the second superstep. As a result, the number of active VMs at the second superstep (6 VMs) is retained for the rest of the supersteps, and this value is smaller than the peak number of VMs used by OPT (8 VMs) which results in a higher cost. 

However, we also see from the core-secs used (Fig.~\ref{fig:coresecs:livj8}) that OPT and FFD use fewer VM cycles than MF/P and LA/P ($112$~core-secs vs. $142$~core-secs), even though this does not reflect in a reduced cost for the VMs used due to the core-min billing granularity. %}makespan constraints for the OPT and FFD algorithms as described earlier.

 % This can partly be seen in the number of core-mins that went unused in these algorithms. Here, the default wasted $14$~core-mins as compared to just $1$~core-min wasted for the OPT, FFD, FFD/P and LA/P algorithms. 
When we consider the data movement costs for OPT-DM, it takes a much longer time to complete compared to the default and other strategies, taking almost $7\times$ longer and also costing more due to this added time. A whole $11$~core-mins of this is under-utilized (Fig.~\ref{fig:underutil:livj8}), due to data movement when the CPU is mostly idle. It can be seen that in the absence of this wastage, the cost would be comparable to OPT and FFD.

For \textbf{USRN/8P}, we see that the makespan for MF/P and LA/P are $6\%$ slower than OPT (Fig.~\ref{fig:makespan:usrn8}). In MF/P and LA/P strategies, all the partitions are pinned to VMs by the third superstep itself. This causes the makespan to increase due to multiple active partitions being on the same VM in future supersteps, and hence increasing the sequentially processing time of those partitions in a superstep. We also see that the cost for LA/P is same as that of MF/P, and those of OPT and FFD due to granualarity of billing. 

The core-secs used by MF/P and LA/P at $227$~core-secs are much higher than $160$~core-secs used by OPT and FFD. But this is not reflected in the core-mins metrics as the MF/P and LA/P algorithms retain a fixed number of VMs after all partitions get pinned, while the OPT and FFD algorithms end up using more VMs to meet the $T_{Min}$ makespan constraint.
For the same reason, the under-utilization, for LA/P and MF/P is more compared to the OPT and FFD.

 %(Fig.~\ref{fig:coremins:usrn8}), and even its core-secs used is close to OPT (Fig.~\ref{fig:coresecs:usrn8}). \note{This is due to the lookahead approach that uses information from future supersteps, this is particularly beneficial for large diameter graphs that can take many supersteps ($14$ supersteps for USRN/8P, compared to $8$ for LIVJ/8P).}\ysnote{RAVI: Why is LA/P much better then FFD/P for USRN? E.g. core-secs is halved.} %  is high, as for USRN, many of the partitions get pinned in later supersteps. As a result, being able to consider the active partitions even in one future superstep can help make better decisions. For e.g., in USRN, it takes \note{$nn$} supersteps to pin all partitions while it only took \note{$mm$} supersteps for LIVJ. As a result, LA/P is able to complete faster than FFD/P and is cheaper as well, getting close to the optimal. 
% The OPT-DM and FFD-FM are slower here as well, taking $7\times$ longer than the default while still being cheaper.

%We also see that the LA/P achieves a smaller wastage of VMs' utilization, and comparable to OPT and $5\times$ smaller than the default (Fig.~\ref{fig:underutil:usrn8}). Despite using half as many core-secs as the default strategy and having a much smaller wasted core-mins, LA/P still costs as much as the default (Fig.~\ref{fig:coremins:usrn8}), and this again is due to the core-min billing granularity. %\ysnote{We need to consider the core-secs too to make a stronger case}

\textbf{ORKT/40P} is a large graph and takes $33$~secs to run using the default strategy (Fig.~\ref{fig:makespan:orkt40}), and with the cost being $40$~core-mins (Fig.~\ref{fig:coremins:orkt40}). We see that OPT and FFD are able to complete at a $40\%$ cheaper cost than the default, saving $16$~core-mins. Here again, LA/P and MF/P are modestly cheaper than the default by $42\%$ while having a small increase in the makespan relative to $T_{Min}$. The core-secs used by LA/P of $854$~core-secs approaches OPT's $667$~core-secs, and is much smaller than the default's $1,330$~core-secs (Fig.~\ref{fig:coresecs:orkt40}). As a result, the under-utilization too is low (Fig.~\ref{fig:coremins:orkt40}), with just $5$~core-min wasted as compared to the default's $13$ wasted core-mins.

% For the smaller number of partitions, FFD/P and LA/P are $30$-$45\%$ slower, respectively, compared to default, while costing less than half the number of core-mins \ysnote{recheck, based on 16P cost. Update explanation.}. 
% For 40 partitions, the makespan of FFD/P and LA/P are even closer to the default, but at the same time, their cost benefit over default is also marginal. Here, we see that the wasted resources of the default is small to begin with, only $12$~core mins, and hence the scope for improvement is lesser. In fact, even OPT and FFD show a cost improvement of only $25\%$ over default, which is the smallest improvement among all cases considered. \ysnote{any other reason why?}
OPT-DM is worse than the default, both in makespan as well as in cost. Given that ORKT is a large graph and each partition has a size of $\sim 100$~MB, the data movement cost for each superstep adds up and causes an increase in both makespan and cost. %\ysnote{why is the cost also higher than default? due to size on disk for orkt?} \drnote{Yes. The avg size of partition for all graphs is >100MB and bandwidth is 16MBps causing atleast 2*6sec overhead added to every superstep which is significantly high }

%\ysnote{One more graph that clearly shows the value of FFD/P or LA/P compared to default?}
In summary, we see that both \emph{MF/P} and \emph{LA/P} are comparable in terms of results, and offer practical strategies to leverage elasticity of VM for partition placement. They are $6$-$29\%$ slower than the default strategy, but $12$-$42\%$ cheaper and consistently use much fewer core-secs to execute than the default. MF/P may be preferable for its simplicity.

%does not have the advantage of LA/P in planning based on a future superstep. Its early pinning of partitions to VMs based on information at the current superstep causes it to under-perform LA/P. It is also occasionally costlier than the default strategy even though the core-secs used by it is smaller than the default in all cases.

The \emph{OPT} and \emph{FFD} strategies guarantee a makespan that matches $T_{Min}$ and are the same or cheaper than the default on cost. They also out-perform all other strategies on core-secs and under-utilization. However, they may not be practical to implement because the data movement costs in unlikely to be ignored. For graphs that are compactly stored, or when if VMs mount the same shared network drive (e.g., AWS's Elastic Block Store (EBS) volumes), these algorithms can be feasible in practice. But if na\"{i}vely transferring data betweens VMs at the end of each superstep, \emph{OPT-DM} is much worse than all other strategies. %\ysnote{For large graphs, the benefits are higher. The VM pricing at core-mins does not show the full benefits of these strategies, and they end up using much fewer core-secs even as the reduction in price paid is smaller.}

% - The improvements in costs are usually higher when the unused utilization of default is higher, since it gives more room for improvement. In cases where the utilization of default is good to begin with, the marginal reduction in costs may not justify the increase in makespan for the FFD/P, LA/P algorithms. 
% - For GoFFish and small IaaS VMs, when data movement of partitions is considered for OPT and FFD, their makespans are much higher than the default and sometimes, they also cost more. 

\section{Conclusions \& Future Work}
\label{sec:conclusion}

In this paper, we have proposed to decouple the partitioning and placement strategies for component-centric graph frameworks to allow flexibility in scheduling them onto elastic VMs that can prevent over-allocation of resources. We have designed several partition placement strategies for graph applications whose runtime behavior can be modeled \emph{a priori}, motivated by our earlier work on meta-graph that can help with coarse-grained static analysis. 

These strategies that include OPT and FFD that are optimal and heuristic formalizations that guarantee the theoretical minimum makespan, as achieved by the default model of over-allocating one VM per partition, while reducing the number of VM used. They also include greedy strategies, MF/P and LA/P, that pin partitions to VMs to avoid data movement, using information on partition timings and VM loads at the current or subsequent supersteps.

The results show that the billing costs, based on realistic IaaS Cloud models, are reduced with marginal increase in makespan for the proposed strategies when evaluated on three different real-world graphs. We also see a more significant improvement in under-utilization and core-secs used, which are orthogonal to the billing granularity.

Leveraging elasticity for distributed graph processing is poorly explored in literature and there are several promising avenues for future work. We propose to examine other non-stationary graph algorithms such as betweenness-centrality and independent-set. This includes the ability to model them using meta-graphs and their benefits from elastic placement. There is scope for using look-ahead heuristics \emph{without pinning} to better leverage the additional knowledge. The optimization goal itself can be modulated to explore the trade-off between makespan and cost.

The minute-granularity is still too coarse for the graphs and application considered, and the strategies may show improved behavior on larger graphs where the makespan is longer, and the core-mins cost much higher. Considering containers such as Docker is also an alternative light-weight ``virtualization'' that can be started and shutdown rapidly, instead of heavy-weight VMs. 

We also plan to examine the impact of the \emph{a priori} predictions being inaccurate -- in our evaluation, we assume perfect knowledge of partition timings but that is unlikely even using the meta-graph model. Here, the static placement strategies proposed here will need to be complemented with dynamic runtime information to update the placements.

%\end{document}  % This is where a 'short' article might terminate
% \ysnote{Extend for IPDPS by having (1) smart placement of partitions to begin with based on meta-graph, that also helps stationary algos. (2) Support for timeseries graphs? (3) Practical comparative validation at scale}
% \ysnote{Use static plus dynamic placement. Statis using meta graph that gives superset of active partitons, dynamic using 1-superstep lookahead decision making}

% \ysnote{Larger graphs with longer execution time will show more benefit. Also, as containerized models get popular, time to start a container is in order of subsecond. Send billing may also move to second based.}

% \ysnote{Use subgraphs a granularity of computation}

% \ysnote{Different VM sizes?}

% \ysnote{Co-planning partitioning and placement. E.g. USRN though it has a large dia is still being pinned early. Use metagraph of partition to move subgraph between partitions?}
% \ysnote{Rune experiments from different source vertices}

% \drnote{In the problem statement, we have specified our primary objective as minimizing the cost. But the optimal solution we mentioned is trying to find out minimum number of VM for each superstep. Also we specify the VM capacity as superstep time. In order to get optimal cost we need to figure out the VM capacity..  }

% \delc{draft}\blindtextc

%ACKNOWLEDGMENTS are optional
%\section{Acknowledgments}

\bibliographystyle{abbrv}
\bibliography{main}  % sigproc.bib is the name of the Bibliography in this case
% ACM needs 'a single self-contained file'!
%
%\balancecolumns % GM June 2007
% That's all folks!
\end{document}